%% file: main.tex
\documentclass[sigconf,screen, nonacm]{acmart}

\input{packages}
\input{macro}
\begin{document}

\title{Are we there yet? An Industrial Viewpoint on \\ Provenance-based Endpoint Detection and Response Tools}


\author{Feng Dong}
\authornote{Hubei Key Laboratory of Distributed System Security, Hubei
Engineering Research Center on Big Data Security, School of Cyber Science and
Engineering, Huazhong University of Science and Technology.}
\affiliation{%
  \institution{Huazhong University of Science and Technology}
  \country{}
  }
\email{dongfeng@hust.edu.cn}

\author{Shaofei Li}
\authornote{Key Laboratory of High-Confidence Software Technologies (MOE), School of Computer Science, Peking University.}
\affiliation{%
  \institution{Peking University}
  \country{}
}
\email{lishaofei@pku.edu.cn}

\author{Peng Jiang}
\authornotemark[2]
\affiliation{%
  \institution{Peking University}
  \country{}
}
\email{pengjiang_pku2020@stu.pku.edu.cn}

\author{Ding Li}
\authornotemark[2]
\authornote{Co-corresponding authors.}
\affiliation{%
  \institution{Peking University}
  \country{}
}
\email{ding_li@pku.edu.cn}

\author{Haoyu Wang}
\affiliation{%
  \institution{Huazhong University of Science and Technology}
  \country{}
  }
\authornotemark[1]
\authornotemark[3]
\email{haoyuwang@hust.edu.cn}

\author{Liangyi Huang}
\affiliation{%
  \institution{Arizona State University}
  \country{}
  }
\email{lhuan139@asu.edu}

\author{Xusheng Xiao}
\affiliation{%
  \institution{Arizona State University}
  \country{}
  }
\email{Xusheng.xiao@asu.edu}

\author{Jiedong Chen}
\affiliation{%
  \institution{Sangfor Technologies Inc.}
  \country{}
  }
\email{chenjiedong1027@gmail.com}

\author{Xiapu Luo}
\affiliation{%
  \institution{The Hong Kong Polytechnic University}
  \country{}
  }
\email{csxluo@comp.polyu.edu.hk}

\author{Yao Guo}
\authornotemark[2]
\affiliation{%
  \institution{Peking University}
  \country{}
}
\email{yaoguo@pku.edu.cn}

\author{Xiangqun Chen}
\authornotemark[2]
\affiliation{%
  \institution{Peking University}
  \country{}
}
\email{cherry@sei.pku.edu.cn}

\renewcommand{\shortauthors}{Dong and Li, et al.}

\begin{abstract}
\ac{pedr} systems are deemed crucial for future APT defenses. Despite the fact that numerous new techniques to improve \ac{pedr} systems have been proposed in academia, it is still unclear whether the industry will adopt \ac{pedr} systems and what improvements the industry desires for \ac{pedr} systems. To this end, we conduct the first set of systematic studies on the effectiveness and the limitations of \ac{pedr} systems. Our study consists of four components: a one-to-one interview, an online questionnaire study, a survey of the relevant literature, and a systematic measurement study. Our research indicates that all industry experts consider \ac{pedr} systems to be more effective than conventional \ac{edr} systems. However, industry experts are concerned about the operating cost of \ac{pedr} systems. In addition, our research reveals three significant gaps between academia and industry: (1) overlooking client-side overhead; (2) imbalanced alarm triage cost and interpretation cost; and (3) excessive server-side memory consumption. This paper's findings provide objective data on the effectiveness of \ac{pedr} systems and how much improvements are needed to adopt \ac{pedr} systems in industry.
\end{abstract}

\maketitle

\input{tex/introduction}
\input{tex/background}

\input{tex/overview}
\input{tex/survey}

\input{tex/questionnaire.tex}

\input{tex/literature}

\input{tex/evaluation}

\input{tex/findings}

\input{tex/discusion.tex}
\input{tex/related}
\input{tex/conclusion.tex}
\balance
\bibliographystyle{ACM-Reference-Format}
\bibliography{ref,xiao}
\clearpage
\input{tex/appendix.tex}
\end{document}

%% file: packages.tex
\usepackage{amsmath,amssymb,amsfonts}
\usepackage{algorithmic}
\usepackage{graphicx}
\usepackage{textcomp}
\usepackage{xcolor}
\usepackage{tcolorbox}

\def\BibTeX{{\rm B\kern-.05em{\sc i\kern-.025em b}\kern-.08em
    T\kern-.1667em\lower.7ex\hbox{E}\kern-.125emX}}

\usepackage[ruled,vlined, noend]{algorithm2e}
\usepackage{mathtools}
\usepackage{url}
\usepackage{booktabs}
\usepackage{multirow}
\usepackage{mdwlist}
\usepackage{xspace}
\usepackage{amsmath}
 \usepackage{amssymb}
\usepackage{misc}
\usepackage{enumerate}

\usepackage{enumitem}
\usepackage{adjustbox}
\usepackage{mdframed}
\usepackage{pifont}
\usepackage{array}
\usepackage{acronym}
\usepackage{wasysym}

\usepackage{graphicx}
\usepackage{float}
\usepackage{subfigure}

%% file: macro.tex
\definecolor{mygreen}{rgb}{0,0.6,0}
\definecolor{mygray}{rgb}{0.5,0.5,0.5}

\newcommand{\eat}[1]{}

\newcommand{\code}[1]{\emph{#1}}

\newif \ifcomments
\commentstrue

\ifcomments
    \newcommand{\kjee}[1]{{-\textcolor{red}{#1}-}}
\else
    \newcommand{\kjee}[1]{}
\fi

\newcommand{\shaofei}[1]{\textsf{\color{green}{({shaofei: #1})}}}
\acrodef{ids}[IDS]{Intrusion Detection System}
\acrodef{edr}[EDR]{Endpoint Detection and Response}
\acrodef{e3}[E3]{Engagement 3}
\acrodef{e5}[E5]{Engagement 5}
\acrodef{apt}[APT]{Advanced Persistent Threats}
\acrodef{pd}[PD]{Provenance Data}
\acrodef{dg}[DG]{Directed Graph}
\acrodef{qg}[QG]{Query Graph}
\acrodef{cti}[CTI]{Cyber Threat Intelligence}
\acrodef{ttp}[TTPs]{Tactics, Techniques, and Procedures}
\acrodef{hsg}[HSG]{High-level Scenario Graph}
\acrodef{tpg}[TPG]{Tactical Provenance Graphs}
\acrodef{soc}[SOCs]{Security Operations Centers}
\acrodef{pedr}[P-EDR]{Provenance-Based Endpoint Detection and Response}
\acrodef{cs}[CS]{Cobalt Strike}
\acrodef{soc}[SOC]{Security Operation Center}

\SetKwInput{KwInput}{Input}
\SetKwInput{KwOutput}{Output}
\SetKwProg{Fn}{Function}{}{end}

\newcommand{\anosec}{AnonymousSec\xspace}
\SetCommentSty{mycommfont}

\usepackage{pifont}


%% file: tex/introduction.tex
\section{Introduction}
\ac{pedr} is a rising next-generation system for APT attack defending~\cite{hossain2017sleuth, hossain2020morse,milajerdi2019holmes,10.1007/978-3-642-40203-6_30,DBLP:journals/corr/ManzoorMVA16,han2020unicorn,wang2020provdetector}. Compared with conventional \ac{edr} systems, \ac{pedr} systems introduce provenance graph, a data structure that models dependencies between system activities, so that they can correlate multiple alarms, leading to higher detection accuracy and better interpretability~\cite{hassan2019nodoze}. 
As such, we have witnessed a rapid growth of \ac{pedr} research in the recent five years from security/system top conferences and industry adoption of \ac{pedr} in commercial products.
According to a recent study~\cite{SoK-History}, there are over 50 \ac{pedr} related papers published in the most prestigious security (IEEE S\&P, CCS, Usenix Security, NDSS) and systems (OSDI, SOSP, ATC) conferences in recent five years. Substantial research efforts have been put forth to improve \ac{pedr} systems in terms of system optimizations~\cite{tang2018nodemerge,hossain2017sleuth,logging2020ccs,valdiating2021ccs}, detection algorithms~\cite{hassan2019nodoze,wang2020provdetector,han2020unicorn,HOLMES,poirot,zengy2022shadewatcher}, and broader security applications~\cite{polinsky2021sciffs,ujcich2021causal}.

While these works have shown promising early results based on evaluations in the academic setting, it is however still unclear whether the industry values the potential of \ac{pedr} systems and would like to adopt some of these works~\cite{SoK-History}.
Moreover, if the industry has not adopted \ac{pedr} systems yet due to various limitations, how these systems can be improved remains unknown.
Knowing the answers to these questions is particularly important, as it can guide future research efforts to focus on the most critical directions based on the industry feedback.
Specifically, there are three key research questions that need to be addressed: 
\begin{itemize}[noitemsep, topsep=1pt, partopsep=1pt, listparindent=\parindent, leftmargin=*]
    \item \textbf{RQ1}: How does the industry compare the effectiveness of \ac{pedr} and conventional \ac{edr}?
    This RQ can help us understand whether the research values of \ac{pedr} systems have been recognized by the industry.
    \item \textbf{RQ2}: What are the bottlenecks for the industry to adopt \ac{edr} systems?
    It is natural that fundamental research takes years before it can be deployed for practical use.
    This RQ can help us focus the efforts in addressing the major bottlenecks and reduce the turnaround time for \ac{pedr} systems to be put into practice.
    \item \textbf{RQ3}: How well can existing \ac{pedr} systems proposed in academia meet the expectations of the industry?
    This RQ can help us understand the gaps between the techniques developed in academia and the expectations of the industry.
\end{itemize}

\begin{figure}[t]
    \includegraphics[width=0.48\textwidth]{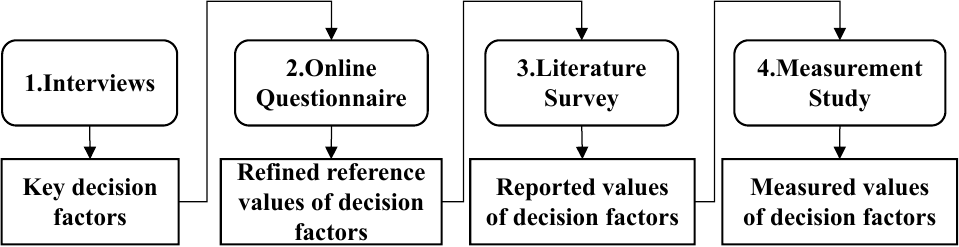}
    \caption{The overview of workflow of our study.}
    \label{fig:survey-workflow}
\end{figure}

To this end, in this paper, we conduct the first set of systematic studies to understand what are the industry's expectations about \ac{pedr} systems and how to close the gaps in adopting \ac{pedr} systems.
More specifically, as shown in Figure~\ref{fig:survey-workflow}, our study consists of four parts:
\begin{itemize}[noitemsep, topsep=1pt, partopsep=1pt, listparindent=\parindent, leftmargin=*]
\item \textit{Interviews}: we first conducted one-to-one interviews to seek feedback on the effectiveness of \ac{pedr} systems and identify their key decision factors in adopting \ac{pedr} systems. 
We successfully recruited ten experienced technical managers of security engineering teams from top IT companies to join our interviews (Section~\ref{sec:interview}).
These companies include both vendors and consumers of \ac{edr}/\ac{pedr} products. 
\item \textit{Online Questionnaire}: based on the key decision factors found in the interviews, we further designed a structured online questionnaire to get feedback from a broader scope of security engineers for refining the reference values of the key decision factors.
Our questionnaire received responses from 48 security engineers in a variety of companies (Section~\ref{sec:questionnaire}). 
\item \textit{Literature Survey}: based on the identified key decision factors, we surveyed the \ac{pedr} systems described in recent publications and evaluated whether they can satisfy these decision factors (Section~\ref{sec:LiteratureSurvey}). 
Our study revealed that none of the existing systems provide evaluation results for all the key decision factors.
\item \textit{Measurement Study}: as many existing systems lack evaluation results for the key decision factors, we further conducted a measurement study on representative \ac{pedr} systems using real industry datasets to measure whether these systems can satisfy these factors and identify how much improvement is needed (Section~\ref{sec:EmpiricalStudy}).
\end{itemize}

We perform an in-depth analysis of the study results and summarize the findings to answer the three research questions:
\begin{itemize}[noitemsep, topsep=1pt, partopsep=1pt, listparindent=\parindent, leftmargin=*]
\item \textbf{RQ1}: All the interviewed managers acknowledged that \ac{pedr} systems are superior than conventional \ac{edr} systems due to better interpretability.
Experienced security analysts can easily understand the provenance data even if it contains only low-level system audit events. 
Surprisingly, while it is natural that fundamental research takes years for it to be deployed in practice, there are already some security teams (2 out of the 10 interviewed teams) that have adopted \ac{pedr} systems.
Furthermore, they have even started to provide training sessions for \ac{pedr} systems. 
These results show that \ac{edr} systems have the potential to replace the  \ac{edr} systems and become the dominating security defense systems for advanced cyber attacks. 

\item \textbf{RQ2}: Most managers considered the operating cost of \ac{pedr} systems, including the computing cost on both the client-side and server-side and the labor cost on alarm triage and attack investigation, as the primary bottleneck in adopting \ac{pedr} systems, even though intuitively we may generally consider detection accuracy as the most important factor.
In fact, most security teams have experiences working with \ac{edr} systems that produce a high number of false positives,s and \ac{pedr} systems generally have higher detection accuracy, and thus they found no problems in using \ac{pedr} systems. 
However, most security teams cannot afford the operating cost of existing \ac{pedr} systems.
For example, provenance data collectors such as Auditd~\cite{auditd} can add at most 821\% more runtime overhead to applications running on the client side,
and some \ac{pedr} systems require more than 200MB memory to process the data for a protected host, which is 10 times more than the industry expectation (20MB/host).
These results show that future research efforts should focus on optimizing the operating cost of \ac{pedr} systems on both the client-side and the server-side.

\item \textbf{RQ3}: By performing a deeper analysis of our study results, we identify three important gaps between the \ac{pedr} techniques proposed by the academia and the expectations of the industry:
\begin{enumerate}[noitemsep, topsep=1pt, partopsep=1pt, listparindent=\parindent, leftmargin=*]
    \item \textbf{Overlooking Client-Side Overhead}: most \ac{pedr} systems (19 out of the 20 surveyed systems) rely on third-party provenance data collectors such as Sysdig~\cite{sysdig} and neglect the client-side overhead.
    \item \textbf{Imbalance between Alarm Triage Cost and Interpretation Cost}: some research focuses on optimizing the precision in reducing alarm triage cost, but it introduces significant interpretation cost by producing a large amount of provenance data to inspect. 
    Similarly, some research focuses on optimizing the interpretation cost but overlooking the precision, producing lots of false positives. 
    Few research has considered both of these factors together, which makes most \ac{pedr} systems impractical in industry settings.
    \item \textbf{Excessive Server-Side Memory Consumption}: most \ac{pedr} systems cache system auditing events in the memory, resulting in very high memory consumption.
    More research efforts are in dire need to optimize memory consumption. 
\end{enumerate}
These identified gaps shed light on what important factors are neglected by the academia and how much improvement of \ac{pedr} systems is needed to meet the industry expectations.

\end{itemize}

In summary, the contributions of this paper are as follows:
\begin{itemize}[noitemsep, topsep=1pt, partopsep=1pt, listparindent=\parindent, leftmargin=*]
    \item We are \textit{the first to investigate the industry's expectations about \ac{pedr} systems} and \textit{provide guidelines on how to close the gaps in adopting \ac{pedr} systems}.
    \item We conduct a one-to-one interview with technical managers from top IT companies and follow up with an online questionnaire to obtain industry expectations on \ac{pedr} systems.
    \item We conduct a measurement study on three representatives \ac{pedr} systems to measure whether existing \ac{pedr} systems meet the industry expectations and how much improvement is needed. We make the dataset and the systems publicly available~\cite{EDREmpiricalStudy} to enable the reproducible study and facilitate further research on APT detection and investigation. 
    \item We perform in-depth data analysis of the study results to identify the gaps between academic techniques for \ac{pedr} systems and the industry expectations and provide guidelines for future research directions.
\end{itemize}





\eat{
rq1
基于我们对采访和问卷的深入分析，我们发现了P-EDR的作用（RQ 1）

rq2
进一步，我们从问卷中归纳出七个工业界EDR决定性因素，其中仅四个必须满足因素。 

rq3
首先，我们选取了xxx
对于文献中没有衡量的指标，我们实现了ABC，并且测量他们在工业环境下数据集下的相关指标值。
随后，我们分析了弥补这些差距存在的挑战，以及相应的未来研究方向。

总之，这篇论文的贡献如下：
1.我们进行了两步调研来调查溯源图技术在工业界应用的作用和主要瓶颈点，并且标准化瓶颈点的度量方法和参考值；
2.我们构建了一个用于APT检测溯源的工业数据集，并且复现了两个经典的P-EDR系统；我们将数据集和复现的系统公开以便于工作的复现和后续APT攻击检测相关研究。
3.我们进行了empirical study来找到PEDR学术界与工业界之间的差距，并且深入分析弥补差距存在的挑战以及后续可能的研究方向。


随着供应链攻击[]、无文件攻击[]等高级攻击手法的流行和应用，APT攻击变得越来越隐蔽。
APT攻击攻破了包括能源公司、政府在内的关键基础目标，造成严重的损失[solarwinds][]。
APT攻击具备高级持续性和隐蔽性的特点，传统的入侵检测工具检测能力有限，大部分还是采用单点检测方法，无法对抗越来越隐蔽的APT攻击。基于溯源图检测的方法，携带丰富的攻击上下文信息，被认为是有希望检测APT攻击的方法。
一些基于图的apt检测方法[holmes,morse, unicorn,provdetector]被提出来用于APT检测，在实验数据集上取得较好的安全效果。然而，基于图的APT检测方法并没有在工业EDR产品中大规模使用。

工业EDR产品中APT检测大部分还是依赖于安全专家人工提取的规则，从system audit log中寻找攻击，部分EDR中的规则支持有限范围日志间关联匹配。基于安全规则的方法安全效果严重依赖安全专家经验，编写成本高且无法检测未知攻击行为，规则无法使用完整的上下文来检测apt复杂的攻击行为。To this end，some graph based methods are applied in EDR  for APT attack forensic. 然而，将所有日志集中到server进行溯源取证的方法，面临极大的传输和存储原始数据的成本，并且检测与溯源过程割裂，无法有效的结合。
It is still unclear about the challenges and pain points of apt detection for industrial endpoints, the understanding of which can be very helpful for combatting the apt attack in industiral settings.
To bridge this gap, we first conduct a comprehensive survey in 大范围的EDR开发者和使用者。首先，我们对5个来自不同leading security company的edr 开发和运营专家进行interviews to gain the big picture to better understand the process of combatting apt attacks in industrial setting. 
采访的结果表明，企业中进行apt检测，xxx，
based on our understanding from the scrum interview， we continue to design a questionnaire to answer RQ1-
问卷调查的结果表明，

1.4measurement to RQ2 and RQ3
观点量化，data支撑

1.5contribution

}

\eat{

把所有的apt文章列全
选取维度
指出研究方向

有亮点的take away points
cost
（1）现有的集中式的apt检测框架成本非常高，成为graph based apt detection方法大规模应用的主要障碍；
现有的方法，面临较高的计算，网络和存储成本；
问卷拿到客观的指标，列个表格,现有方法是否满足
制约的因素，主要是成本太高，告警数量多延迟高，图太大无法研判，图语义gap

传统的detection results，

攻击研判需要图吗? 究竟需要什么信息来研判,图有没有用,如果没用需要在哪些方面改进

攻击发生到检出时间；

研究方向，不是准确率

公司的人员，

成本是第一位的

准确率是第二位

图构建和检测，需要消耗大量的内存（300-600MB/host），成为大规模集群应用的主要障碍之一；
现有的方法如果保护1000台主机规模集群，需要重新构建几乎60\%同等规模硬件集群

mean time to detect

balance of precision and recall
（2）现有方法面临比较大的误报问题，需要想办法控制误报水平；

除了recall precision 还有告警的绝对数量
重复告警，碎片化告警

现有的方法需要在最佳的recall条件下，将alarm rate控制在安全分析人员可以处理的范围（10 alarms/day/host）内；

（3）检测结果的粒度非常重要，graph level和path level的结果，需要进一步的指出哪些节点；

graph level的检测无法将检测结果节点规模控制到200范围以内；
\shaofei{响应流程, 如何处理告警}

（4）semantic gap
1. 能够提炼high-level的语义很好，但是没有也可以用，不是必须的
2. 实现自动的语义翻译比较困难，现有的不好用，需要人工经验

输入输出，输出的图怎么怎么样，对比

overall
目前，没有一种方法能够很好的满足上述的三种方法；
开发一套工业环境下使用的apt检测工具，需要面临以下技术挑战：
1.如何降低开发成本；
2.如何控制告警率；
3.如何让检测结果可响应；

进而引出研究方向

}

\eat{
7.3 讨论，本文的三个核心论点：
1. 成本：现有的apt检测框架成本非常高，graph based apt detection方法大规模应用的主要成本问题是内存问题，与图大小息息相关
2. 衡量指标：只有precision，recall的衡量是不够的，还要有告警绝对数量，告警粒度等指标，保证后续人可处理。
3. 语义gap：能够提炼high-level的语义很好，但是没有也可以用，不是必须的，实现自动的语义翻译比较困难，现有的不好用，需要专业的人工经验。
}

%% file: tex/background.tex
\section{Background}
In recent years, research on provenance analysis is emerging in academia, and it has gradually become an effective tool for \ac{apt} detection. Muhammad~\cite{SoK-History} describes provenance analysis as the totality of system execution and facilitates causal analysis of system activities by reconstructing the chain of events that lead to an attack as well as the ramifications of the attack. BackTracker~\cite{king2003backtracking} identifies files and processes that may affect the detection point and displays the chain of events in a provenance graph, which is the first attempt on provenance-based intrusion detection. Due to provenance auditing can record system activities in detail and is hard to evade, provenance-based \ac{apt} detection models~\cite{hossain2017sleuth,hossain2020morse,milajerdi2019holmes,hassan2020rapsheetl,DBLP:journals/corr/ManzoorMVA16,han2020unicorn,wang2020provdetector,9833669} have emerged in the past few years. However, according to our survey, provenance-based techniques have not been widely used in industrial commercial \ac{edr}. There are still unacceptable gaps between academic research and industrial deployment.

\subsection{Overview of the \ac{pedr} System}
The overall process of a \ac{pedr} system is shown in Figure~\ref{fig:EDRworkflow}. In general, a \ac{pedr} system is the core part of a commercial \ac{soc} that monitors the endpoint hosts (e.g., servers, desktops, laptops, e.t.c.) and detects attacks on the hosts. A typical \ac{pedr} system consists of two key components: the client-side component and the server-side component. The client-side component is an agent installed on the monitored hosts that collects provenance data from the hosts. 
The server-side component is a dedicated server that processes the collected provenance data and detects APT attacks. 
A typical \ac{pedr} system~\cite{hassan2019nodoze,HOLMES,han2020unicorn,hassan2020rapsheetl} contains four key steps. 

The first step is data collection, which runs on the monitored hosts to collect provenance data and do some preliminary refinement and cleaning. Normally, the collected provenance data contains process, file, register, and network operation logs. Then, the \ac{pedr} system sends the collected data to the server. In commercial systems, the agent may also compress the provenance data before sending it to the server. 

The next three steps are on the server side. The second step is detection, in which the \ac{pedr} system detects APT attacks from the collected provenance data, using manually crafted rules~\cite{HOLMES,hassan2020tactical} or machine learning algorithms~\cite{wang2020you,han2020unicorn}. 
The third step is the investigation, in which the \ac{pedr} automatically helps security admins correlate related alarms and investigate the root causes of alarms. 
In the last step, security experts validate the generated alarms and respond to possible attacks. 

\subsection{Provenance Analysis and Provenance Graph}
Compared with conventional \ac{edr} systems, the unique advantage of a \ac{pedr} system is that it automatically reconstructs the dependencies between log entries and alarms in the step of investigation~\cite{hassan2019nodoze,HOLMES}. The alarms of conventional \ac{edr} systems are isolated. Thus, it is particularly hard for security admins to combine related alarms or recover their root causes. On the flip side, \ac{pedr} systems use provenance graphs to model the data and control dependencies between events in provenance data, automatically linking related alarms and their root causes together, leading to more interpretable detection results.

In \ac{pedr} systems~\cite{277080,wang2020you,HOLMES,9152772,sleuth,263852,zeng2021watson}, a provenance graph is a directed graph constructed from \textit{system auditing events}, where each event represents a system activity.
Formally, system auditing events are represented as three-tuples $\langle$subject, operation, object$\rangle$. The subject and the object represent \textit{system entities}, and the operation represents an action performed by the subject on the object. The typical values for the three-tuple are shown in Table~\ref{tab:system-call}, in which $\leftrightarrow$ means the entities on both sides can be subjects or objects. In a provenance graph, the nodes are system entities, and the edges are the actions. The directions of edges represent the dependencies of data or control flow. 

\subsection{Example Provenance Analysis}
\begin{figure}[t!]
    \setlength{\abovecaptionskip}{5pt}
    \includegraphics[width=0.48\textwidth]{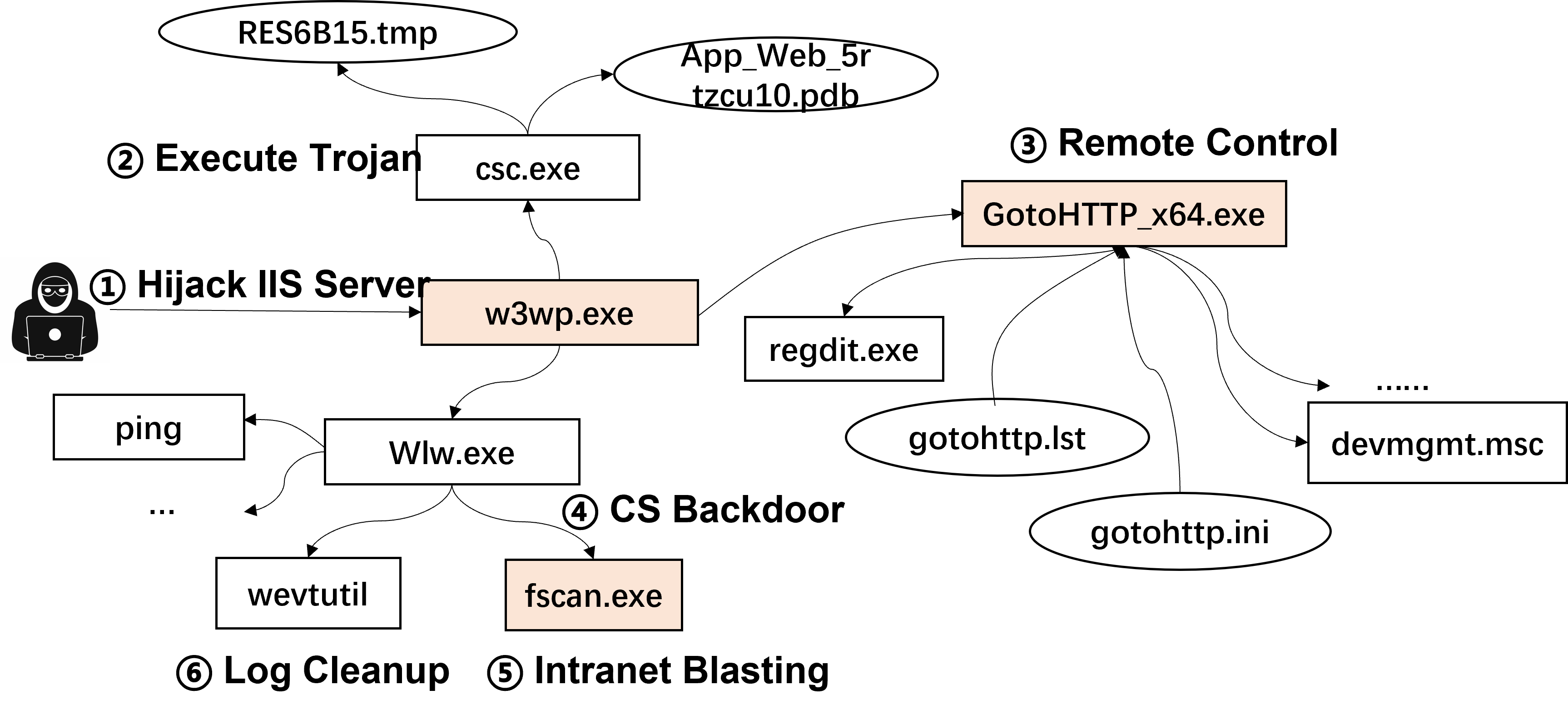}
    \caption{An example of an APT attack that hijacks Windows IIS Web server and leaves a \ac{cs} backdoor for lateral movement. Then it uploads the remote control tool for privilege escalation.}
    \label{fig:apt-example}
\end{figure}

In Figure~\ref{fig:apt-example}, we show an example of the provenance graph for a real APT attack. In this attack, the adversary first hijacks the Windows IIS Web Server ``w3wp.exe'' through a web shell. Then she uses ``csc.exe'' to execute a trojan. The adversary also runs the remote tools ``GotoHTTP\_x64.exe'' to modify the registry privilege escalation. Lastly, she leaves a backdoor ``Wlw.exe'' for intranet blasting with ``fscan.exe'' and uses ``wevtutil'' to clear footprints. The orange nodes are alarms generated by the detection system. 

In this example, we notice that the provenance graph links multiple alarms based on their dependencies. It also backtracks the entry of the attacks so that security admins can recover the root causes of alarms. Therefore, security analysts consider \ac{pedr} systems more accurate and intuitive for APT attack detection and investigation, leading to the popularity in academia~\cite{277080,wang2020you,HOLMES,9152772,sleuth,263852,zeng2021watson}. 

\begin{table}[!t]
    \setlength{\abovecaptionskip}{5pt}
    \centering
    \caption{System events of the provenance analysis}
    \begin{tabular}{c c}
    \hline
    \textbf{Entity$\leftrightarrow$Entity} & \textbf{Operation Types} \\ \hline
    Process$\rightarrow$File & read, write, create, chmod, rename\\
    Process$\leftrightarrow$Process & fork, clone, execve, pipe \\
    Process$\rightarrow$IP & sendto, recvfrom, recvmsg, sendmsg \\
    \hline
    \end{tabular}

    \label{tab:system-call}
\end{table}


\begin{figure*}[!ht]
    \setlength{\abovecaptionskip}{5pt}
    \centering
    \includegraphics[width=0.99\textwidth]{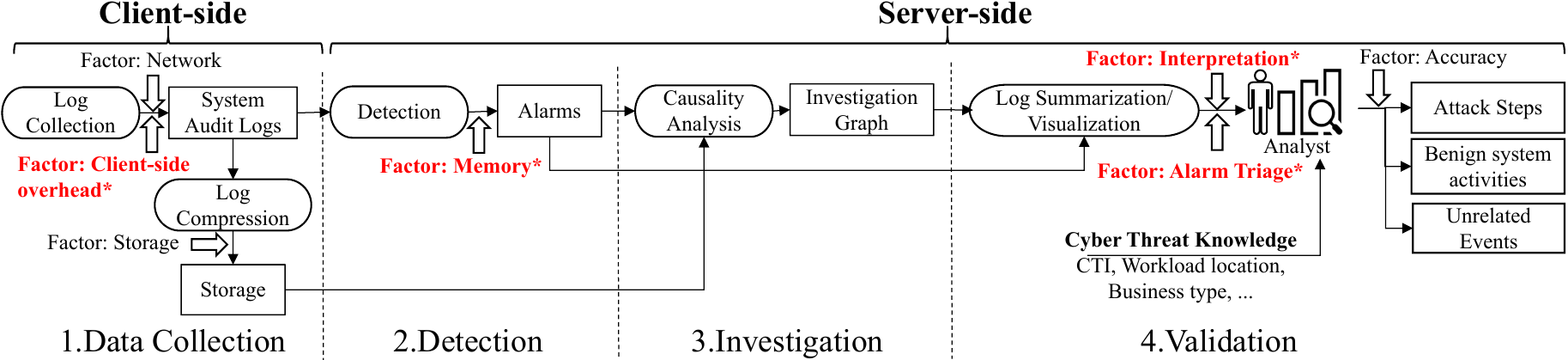}
    \caption{The workflow of using \ac{pedr} to detect attacks }
    \label{fig:EDRworkflow}
    \end{figure*}


\subsection{Ethical Consideration of This Work}
This work was approved by our institution, and we strictly follow our institution's research data management policy, including data storage, sharing, and disposal. 
The data collected from the participants in the interviews and questionnaires were carefully processed. In both the interview and the online questionnaire, we acquired the consent of the participants and confirmed that our interviews accurately reflected their own opinions.

%% file: tex/overview.tex
\eat{

\section{Overview}
\label{sec:overview}
To reveal the gap of \ac{pedr} between academic and industry, we conduct a systematic study that consists of four parts, the workflow of which is summarized in Figure~\ref{fig:survey-workflow}. 
We first conducted a one-to-one interview with ten experts of security teams from top IT companies to seek feedback on the effectiveness of \ac{pedr} systems and identify their top decision factors in adopting \ac{pedr} systems. Then we obtained the accurate reference values of the must-meet factors through an online questionnaire based on the results of the interview. Third, we surveyed the \ac{pedr} systems in recent publications and evaluated whether they can satisfy the must-meet factors from the industry. At last, for the factors that are not measured in the surveyed papers, we conducted experiments in measurement study.

\begin{figure}[t!]
    \setlength{\abovecaptionskip}{5pt}
    \includegraphics[width=0.48\textwidth]{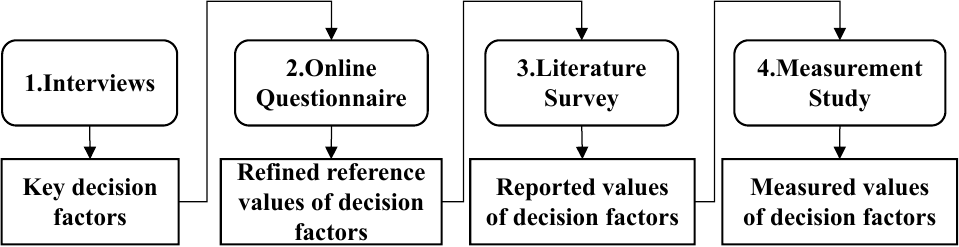}
    \caption{The overview of workflow of our study.}
    \label{fig:survey-workflow}
\end{figure}
}

%% file: tex/survey.tex


\eat{
首先，我们对行业中的专家进行采访，得到当前工业界环境下进行APT检测需要考虑的关键因素。随后，基于这些关键因素，我们设计了结构化的问卷,并且将问卷分发给不同SOC的EDR研发和分析人员。问卷主要设计来量化工业界环境下APT检测系统需要满足的关键因素，进而衡量现有APT检测方法应用于工业主机的差距。

motivation.x provenance graph 使用率低
gap

RQ1 Effectiveness in industry adoption:
1. do people watch provenance graph? 40\% see. intention to use provenance: yes. current situation for preferring proveneance over EDR: no. why? In 2 in use, all use provenance. 
-> very promising. Reason: "a few sentences"
provenance graph中的上下文比单条日志，包含更多的攻击信息。对apt拥有更好的检测和溯源能力。

2. \textbf{attack/ application level/ sys logs do not matter much}, semantics can be understood without difficulties. experience is very important.
companies usually have corresponding training. cyber education is lacking but is improving. cyber education is much needed future development. zoom.
-> ?
将低级别的系统日志翻译到高级别的行为，与分析师的经验相关。告警语义不是一个严重的问题，

3. other improvements?
-> frequent lightweight update, major version update cannot be too frequent
-> must be able to preserve the current functioning of the clients

RQ2 Bottlenecks in practical usage:
1. cost vs precision 
overall cost = machine cost + triage cost + investigation cost
-> 
2. labor cost 1: often neglected = triage cost + investigation cost
 labor cost 2: trivial attacks, cannot distinguish these, number of alarms, warning/critical
->

3. machine cost 1: central architecture vs distributed architecture: client-optimized architecture is preferred, i.e., agent cost is a major factor
->
5. machine cost 2: server memory
}

\section{One-To-One Interviews}
\label{sec:interview}

To seek feedback on the effectiveness of \ac{pedr} systems and identify the decision factors for the adoption of \ac{pedr} systems in the industry, we conducted one-on-one interviews with experienced technical managers from top IT companies.
We next describe the participants, the interview methodology, and the result.

\input{tab/interview_participants.tex}


\subsection{Participant Recruitment}
We recruited participants from \ac{edr} developers and consumers, who have the first-hand experiences of \ac{edr} in the industry. 
We chose 6 \ac{edr} vendors from top-tier endpoint security companies, and 6 consumers of \ac{edr} systems from diverse kinds of organizations, including IT, education, transportation, and manufacturing.
The consumers of \ac{edr} and \ac{pedr} systems include ByteDance (the world-leading social media provider), MeiTuan (one of the leading AI companies in China), Peking University (one of the most famous universities in China), S.F Express (the biggest express company in China), and FiberHome (the famous manufacturer for IoT devices). 
The vendors of \ac{edr} and \ac{pedr} systems are among the top security vendors in China~\cite{ChinaCompaniesRank} and the world~\cite{CybersecurityCompaniesRank}, including Tencent Security~\cite{TencentSecurity}, Trend Micro~\cite{TrendMicro}, Sangfor~\cite{sangfor}, Rising~\cite{Rising}, and NSFOCUS~\cite{nsfocus}. 

We first found the points of contact (POC) of EDR through the company website, social media, and product technical support list, and these POCs recommended 12 technical managers. 
Ten managers (average of 10+ years of experience)  agreed to participate in our interview.

\noindent\textbf{Participant Background}:
Our participants are experienced leaders in security. They have, on average, 10.5 years of experience, ranging from 5-21 years. 
Each of them leads a technical team with 25-30 engineers on average. 
Our participants are all very familiar with provenance analysis techniques and \ac{pedr} systems.  
Specifically, $E1$ and $E2$ are already using \ac{pedr} systems in their companies, and $E6$ and $E7$ are the developers of the \ac{pedr} systems. 
$E3$, $E4$, and $E5$ who are not using \ac{pedr} are familiar with provenance analysis techniques and are considering using these techniques in the future. 
Lastly, the remaining three ($E8$, $E9$, and $E10$) who are not developing provenance analysis techniques in their current products are also very knowledgeable about the recent progress in academia and may adopt \ac{pedr} when it is necessary. 
Table~\ref{tab:Interview_backgroud} shows the detailed background information of the participants.

\subsection{Interview Methodology}
We interviewed each manager via a 30-min online video conference. 
All managers chose to participate in our study voluntarily as they expect our research results can better help them develop and use \ac{edr}/\ac{pedr} systems.
To ensure the objectiveness of our interview, we followed the principles in \textit{Qualitative Interview Design}~\cite{turner2022qualitative,mann2016research}. Specifically, we explained the purpose of our interview before the interviews and told them how to get in touch with us later if they want to. 
We designed all the interview questions to be open-ended, and the participants are able to choose their own terms when answering questions. 
We also designed our questions to avoid words that might influence answers. 

\noindent\textbf{Interview Questions}:
Our interview questions consist of two parts. The first part is the background session, where we ask the participants to introduce their technical backgrounds, including organization name, job title, years of experience, team size, and experiences with \ac{pedr} systems. 
The second part of our interview questions is the opinion session, which includes questions about the participants' opinions on the key decision factors and the limitations of \ac{edr} and \ac{pedr} systems. 
Below are our interview questions:
\begin{itemize}[noitemsep, topsep=1pt, partopsep=1pt, listparindent=\parindent, leftmargin=*]
    \item Do you think \ac{pedr} systems are more effective than conventional \ac{edr} systems?
    \item What are the limitations of existing \ac{edr}/\ac{pedr} products?
    \item What are the key decision factors when you decided to adopt your current \ac{edr}/\ac{pedr} solution? Are these factors must-meet or optional? Can you rank these factors? 
    \item What metrics do you use to measure these key decision factors?

\end{itemize}



Besides the background session and the opinion session, we also asked the participants several casual questions to help them relax. 
These questions are not related to our research objective but facilitate the participants to share their true opinions~\cite{mann2016research}. 
\input{tab/interview_effectiveness.tex}

\noindent\textbf{Data Processing:} With the authorization of the interviewees, their identities were anonymized, and their interviews were saved in audio form. 
We transcribed the audio into text using a popular Speech-to-Text conversion tool. 
Two authors independently inspected the converted texts and cross-checked the results. We sent the verified texts back to the interviewees for confirmation. 
The data retention period is one year, and we will get further authorizations from the interviewees when the retention period expires.

\subsection{Results}
\label{subsec:interview_result}
\noindent\textbf{Effectiveness of \ac{pedr}:} Table~\ref{tab:Interview_effectiveness} summarizes the answers of our participants regarding the effectiveness of \ac{pedr}.
Overall, all of them agree that \ac{pedr} systems are more effective than conventional \ac{edr} systems, and four managers have already adopted \ac{pedr} systems. 
We will discuss the details of their opinions in Section~\ref{sec:findings}. 

\noindent\textbf{Key Decision Factors:}
We have summarized seven key decision factors mentioned by the participants in 
Table~\ref{tab:FactorDefinitions0}, including \textsf{Network}, \textsf{Storage}, \textsf{Memory}, \textsf{Client-Side Overhead}, \textsf{Interpretation}, \textsf{Alarm Triage}, and \textsf{Accuracy}.
Particularly, \textsf{Accuracy} represents the \textit{detection effectiveness} of a \ac{pedr} system, while the other factors represent the \textit{operating cost} of a \ac{pedr} system. Thus, we further classified them into three major categories: ``Computing Cost'' (\textsf{Network}, \textsf{Storage}, \textsf{Memory}, \textsf{Client-Side Overhead}), ``Labor Cost'' (\textsf{Interpretation} and \textsf{Alarm Triage}), and ``Performance'' (\textsf{Accuracy}). 
Among these factors, we have identified four must-meet factors (i.e., highlighted by all the participants who mentioned such factors), including \textsf{Memory}, \textsf{Client-Side Overhead}, \textsf{Interpretation}, and \textsf{Alarm Triage}. For clarity's sake, we present these must-meet factors in the workflow of \ac{pedr} (see Fig.~\ref{fig:EDRworkflow}).

Further, we summarize the expected values for these key decision factors provided by each participant in Table~\ref{tab:InterviewResult}.
The last row of Table~\ref{tab:InterviewResult} shows the reference ranges for each decision factor.  
The lower bound and the higher bound of each reference range are the minimum and maximum estimated values provided by our participants, respectively. 
We next depict each of them.

\noindent\textbf{Computing Cost:}
For the computing cost, the participants have expressed concerns about the average memory consumption on the server side (\texttt{ServerMem}). On average, they expect the server to consume less than 27.6MB of memory per monitored host.

The developers of \ac{edr} systems ($E6$, $E7$, $E8$, $E9$ and $E10$) consider network cost and storage cost as optional decision factors. In other words, the developers agreed that the network and storage costs were important, but they were acceptable if a \ac{pedr} system did not meet the requirements. With respect to the metrics, the developers agreed to use the average bandwidth utilization of a \ac{pedr} and the average disk utilization to measure the network cost and storage cost, respectively. 
The consumers of \ac{edr} systems ($E1$, $E2$, $E4$, and $E5$) did not mention the requirements for network and storage costs, except for $E3$. 
$E3$ requires the storage cost should not exceed 10\% of the total disk size of the monitored systems.

For the client-side overhead, the participants believe that two metrics are useful.
The first metric is the average runtime overhead on the monitored machine (\texttt{RT OH}), and the second one is the average memory consumption on the monitored machine (\texttt{ClientMem}). On average, the participants expect a \ac{pedr} system introduces a performance overhead of less than 5.2\% and consumes less than 160MB of memory on each monitored host.

\noindent\textbf{Labor Cost:}
\label{sec:laborcost}
Labor cost lies in the interpretation and alarm triage.
For the cost of interpretation, the four participants ($E1$, $E2$, $E6$, and $E7$) who are either using provenance analysis techniques or developing provenance-analysis-based solutions also marked the interpretation cost as a must-meet factor. 
They agreed to use the average number of nodes of provenance graphs of alarms as the metric for the interpretation cost. 
The reason is that \ac{pedr} systems generate a provenance graph for each alarm to reveal its context. 
Thus, the size of the provenance graphs determines the workload for the security team to interpret the alarms.
Particularly, the participants expect the number of nodes in provenance graphs to be between 10 to 100.   

For alarm triage, the participants agreed to use the average number of alarms per monitored host per day to measure the triage cost. 
Even though precision is directly related to the triage cost, the average number of alarms per monitored host per day is more intuitive for cost estimation since it is positively correlated with the number of alarms. 
8 out of the 10 participants mentioned that their teams or customers have a fixed number of analysts to investigate the alarms, and thus they can only process a limited number of alarms per day. 
The expected average number of alarms per host per day ranges from $0.001$ to $0.1$. 

\noindent\textbf{Performance:}
Only two participants ($E2$ and $E3$) considered accuracy as one of the decision factors for choosing \ac{edr}/\ac{pedr} systems. 
Others argue that while accuracy is important, accuracy-related issues can be resolved by upgrading security rules or models within a reasonable time. 
In terms of the metrics, both $E2$ and $E3$ agreed to use precision to measure the accuracy. 
The reason is that, in practice, the recall and other metrics are difficult to evaluate due to the lack of ground truth. 
The expected value for precision ranges from $0.85$ to $0.9$. 
Note that the participants acknowledged the importance of precision, but they preferred to use the average number of alarms per monitored host per day to evaluate the performance of a \ac{pedr} system.

\input{tab/key_factors.tex}
\input{tab/interview_result.tex}

%% file: tab/interview_participants.tex
\begin{table*}
\centering
\setlength{\abovecaptionskip}{5pt}
\caption{Background information of the interview participants}
\label{tab:Interview_backgroud}
\resizebox{0.99\textwidth}{!}{\begin{tabular}{|r|l|l|l|l|r|r|l|} 
\hline
\textbf{ID}  & \textbf{Role}                      & \textbf{Company Name}      & \textbf{Industry Are}a & 
\multicolumn{1}{c|}{\textbf{Job Title}}                               & \textbf{Years of Exp.} & \textbf{Team Size}  & \textbf{Adopt \ac{pedr}}  \\ 
\hline
E1  & \multirow{5}{*}{Consumer} & ByteDance         & Technology    & Head of Server Security                 & 6             & 20$\sim$25 & Yes       \\ 
\cline{1-1}\cline{3-8}
E2  &                           & MeiTuan           & Technology    & Cloud Workload Security Leader          & 5             & 20$\sim$25 & Yes       \\ 
\cline{1-1}\cline{3-8}
E3  &                           & Peking University & Education     & Director of Network Security Office     & 19            & 10$\sim$15 & No      \\ 
\cline{1-1}\cline{3-8}
E4  &                           & S.F. Express      & Transportation       & Endpoint Security Manager               & 10            & 20$\sim$25 & No      \\ 
\cline{1-1}\cline{3-8}
E5  &                           & FiberHome         & Manufacturing & Endpoint Security Manager               & 8             & 5$\sim$10  & No      \\ 
\hline
E6  & \multirow{5}{*}{Vendor}   & Tencent Security       & Security      & Director of EDR                         & 10            & 10$\sim$15 & Yes    \\ 
\cline{1-1}\cline{3-8}
E7  &                           & Trend Micro        & Security      & Detection Engine Architect of EDR       & 9             & 20$\sim$25 & Yes    \\ 
\cline{1-1}\cline{3-8}
E8  &                           & Sangfor           & Security      & Director of Workload Protection Product & 8             & 65$\sim$70 & No      \\ 
\cline{1-1}\cline{3-8}
E9  &                           & Rising            & Security      & EDR Architect                           & 21            & 50$\sim$55 & No      \\ 
\cline{1-1}\cline{3-8}
E10 &                           & NSFOCUS           & Security      & EDR Product Manager                     & 9             & 30$\sim$35 & No      \\
\hline
\end{tabular}}
\end{table*}

\eat{
\begin{table*}
\centering
\caption{Background information of the interview participants}
\label{tab:Interview_backgroud}
\resizebox{0.99\textwidth}{!}{\begin{tabular}{|r|l|l|l|l|r|r|} 
\hline
\textbf{ID}  & \textbf{Role}                      & \textbf{Company Name}      & \textbf{Industry Area} & 
\multicolumn{1}{c|}{\textbf{Job Title}}                               & \textbf{Years of Exp.} & \textbf{Team Size}   \\ 
\hline
E1  & \multirow{5}{*}{Consumer} & ByteDance         & Technology    & Head of Server Security                 & 6             & 20$\sim$25       \\ 
\cline{1-1}\cline{3-7}
E2  &                           & MeiTuan           & Technology    & Cloud Workload Security Leader          & 5             & 20$\sim$25        \\ 
\cline{1-1}\cline{3-7}
E3  &                           & Peking University & Education     & Director of Network Security Office     & 19            & 10$\sim$15       \\ 
\cline{1-1}\cline{3-7}
E4  &                           & S.F. Express      & Transportation       & Endpoint Security Manager               & 10            & 20$\sim$25       \\ 
\cline{1-1}\cline{3-7}
E5  &                           & FiberHome         & Manufacturing & Endpoint Security Manager               & 8             & 5$\sim$10        \\ 
\hline
E6  & \multirow{5}{*}{Vendor}   & Tencent Security       & Security      & Director of EDR                         & 10            & 10$\sim$15     \\ 
\cline{1-1}\cline{3-7}
E7  &                           & TrendMicro        & Security      & Detection Engine Architect of EDR       & 9             & 20$\sim$25     \\ 
\cline{1-1}\cline{3-7}
E8  &                           & Sangfor           & Security      & Director of Workload Protection Product & 8             & 65$\sim$70       \\ 
\cline{1-1}\cline{3-7}
E9  &                           & Rising            & Security      & EDR Architect                           & 21            & 50$\sim$55       \\ 
\cline{1-1}\cline{3-7}
E10 &                           & NSFOCUS           & Security      & EDR Product Manager                     & 9             & 30$\sim$35       \\
\hline
\end{tabular}}
\end{table*}
}

%% file: tab/interview_effectiveness.tex
 \begin{table}[]
    \setlength{\abovecaptionskip}{5pt}
     \caption{Interview results for \ac{pedr} effectiveness}
     \label{tab:Interview_effectiveness}
     \resizebox{0.49\textwidth}{!}{
        \begin{tabular}{|ll|}
        \hline
        \multicolumn{1}{|l|}{\textbf{Answers}}       & \textbf{Participants}                   \\ \hline
        \multicolumn{2}{|l|}{\textbf{Limitations of  \ac{edr}/\ac{pedr}}}                                              \\ \hline
        \multicolumn{1}{|l|}{High Client-Side Overhead}      & E1, E2, E3, E4, E5, E6, E7, E8, E9, E10 \\ \hline
        \multicolumn{1}{|l|}{Too Many False Alarms}             & E1, E2, E4, E5, E6, E7, E8              \\ \hline
        \multicolumn{1}{|l|}{Incomplete Rule Set}          & E1, E2, E4, E5, E7, E9, E10             \\ \hline
        \multicolumn{1}{|l|}{Data Privacy}            & E3                                      \\ \hline
        \multicolumn{2}{|l|}{\textbf{Effectiveness of \ac{pedr}}}                      \\ \hline
        \multicolumn{1}{|l|}{\ac{pedr} Already Deployed} & E1, E2, E6, E7                          \\ \hline
        \multicolumn{1}{|l|}{\ac{pedr} Better Than \ac{edr}} & E1, E2, E3, E4, E5, E6, E7, E8, E9, E10                     \\ \hline
        \end{tabular}
     }
 \end{table}

 \eat{
  \begin{table}[]
     \caption{Limitations and Effectiveness as Reported by participants}
     \resizebox{0.5\textwidth}{!}{
     \begin{tabular}{|ll|}
     \hline
     \multicolumn{1}{|l|}{\textbf{Features}}       & \textbf{participants}                   \\ \hline
      \multicolumn{2}{|l|}{\textbf{Limitations}}                                              \\ \hline
     \multicolumn{1}{|l|}{High Client-side Overhead}      & E1, E2, E3, E4, E5, E6, E7, E8, E9, E10 \\ \hline
     \multicolumn{1}{|l|}{Too Many False Alerts}             & E1, E2, E4, E5, E6, E7, E8              \\ \hline
    \multicolumn{1}{|l|}{Incomplete Rule Set}          & E1, E2, E4, E5, E7, E9, E10             \\ \hline
    \multicolumn{1}{|l|}{Data privacy}            & E3                                      \\ \hline
     \multicolumn{2}{|l|}{\textbf{Effectiveness}}                      \\ \hline
     \multicolumn{1}{|l|}{PAT used in current EDR} & E1, E2, E6, E7                          \\ \hline
    \multicolumn{1}{|l|}{interested in using PAT} & E1, E2, E3, E4, E5, E6, E7, E8, E9, E10                     \\ \hline
     \end{tabular}
     }
     \label{tab:Interview_effectiveness}
 \end{table}
 }

%% file: tab/key_factors.tex
\begin{table}
    \setlength{\abovecaptionskip}{5pt}
    \centering
    \caption{Definitions of the Seven Key Factors}
    \label{tab:FactorDefinitions0}
    \resizebox{0.49\textwidth}{!}{\begin{tabular}{|l|p{150pt}|} 
    \hline
    \textbf{Factor}     & \textbf{Description}                                                                                                   \\ 
    \hline
    \multicolumn{2}{|l|}{\textbf{Computing Cost}}                                                                                                  \\ 
    \hline
    \multirow{2}{*}{\begin{tabular}[c]{@{}l@{}}CC1: \textsf{Client-Side Overhead} \end{tabular} }
          & how much an \ac{edr} system slows down the protected hosts  \\ 
    \hline
    \multirow{2}{*}{CC2: \textsf{Network}}        & bandwidth occupied by transmitting system audit logs to the server                \\ 
    \hline
    CC3: \textsf{Storage}      & hard-disk used to store the system logs                                    \\ 
    \hline
    \multirow{2}{*}{CC4: \textsf{Memory}}      & server memory size required to analyze the collected logs                 \\ 
    \hline
    \multicolumn{2}{|l|}{\textbf{Labor Cost}}                                                                                                    \\ 
    \hline
    LC1: \textsf{Alarm Triage}        & man-hour required to detect false alarms                                     \\ 
    \hline
    \multirow{2}{*}{LC2: \textsf{Interpretation}} & man-hour required to interpret attack results                               \\ 
    \hline
    \multicolumn{2}{|l|}{\textbf{Performance}}                                                                                                   \\ 
    \hline
    \textsf{Accuracy}       & attack detection accuracy                                                                                           \\
    \hline
    \end{tabular}}
    \end{table}

%% file: tab/interview_result.tex
\begin{table*}[t!]
    \setlength{\abovecaptionskip}{5pt}
    \centering
    \caption{Interview results for key decision factors. 
    Each cell for a factor shows a rank of importance provided by the corresponding participant followed by a list of metric values to evaluate the factor. Symbol $*$ indicates the must-meet factor.
    }
    \label{tab:InterviewResult}
    \resizebox{1\textwidth}{!}{\begin{tabular}{|r|r|r|r|r|r|r|r|} 
    \hline
                                                              & \multicolumn{4}{c|}{\textbf{Computing Cost}}                                                                                                                                                                                                                                                             & \multicolumn{2}{c|}{\textbf{Labor Cost}}                                                                                                       & \textbf{Performance}                                           \\ 
    \hline
    \textbf{ID}                                               & \textbf{Network}                                                     & \textbf{Storage}                                                      & \textbf{Memory*}                                                  & \textbf{Client-Side Overhead*}                                                     & \textbf{Interpretation Cost* \textbf{~ ~}}                           & \textbf{Alarm Triage Cost* \textbf{\textbf{\textbf{~ ~}}}}                & \textbf{Accuracy}                                              \\ 
    \hline
    E1                                                        & None                                                                 & None                                                                  & \begin{tabular}[r]{@{}r@{}}3, ServerMem*: \\ 30MB/host\end{tabular}  & \begin{tabular}[r]{@{}r@{}}2, ClientMem*: 100MB/host, \\ RT OH*:1\%\end{tabular}  & \begin{tabular}[r]{@{}r@{}}4, Number of \\ nodes*: 100\end{tabular} & \begin{tabular}[r]{@{}r@{}}1, Alarms*:\\ 0.001/day/host\end{tabular}  & None                                                           \\ 
    \hline
    E2                                                        & None                                                                 & None                                                                  & \begin{tabular}[r]{@{}r@{}}3, ServerMem*: \\ 50MB/host\end{tabular}  & \begin{tabular}[r]{@{}r@{}}1, ClientMem*: 150MB/host, \\ RT OH*:5\%\end{tabular}  & \begin{tabular}[r]{@{}r@{}}4, Number of \\ nodes*: 10\end{tabular}  & \begin{tabular}[r]{@{}r@{}}2, Alarms*:\\ 0.001/day/host\end{tabular}  & \begin{tabular}[r]{@{}r@{}}5, Precision, \\ > 0.85\end{tabular}  \\ 
    \hline
    E3                                                        & None                                                                 & \begin{tabular}[r]{@{}r@{}}3, Disk: \\60MB/day/host\end{tabular}   & \begin{tabular}[r]{@{}r@{}}2, ServerMem*: \\ 30MB/host,\end{tabular} & \begin{tabular}[r]{@{}r@{}}1, ClientMem*: 100MB/host, \\ RT OH*:5\%\end{tabular}  & None                                                                & None                                                                     & \begin{tabular}[r]{@{}r@{}}5, Precision, \\ > 0.9\end{tabular}   \\ 
    \hline
    E4                                                        & None                                                                 & None                                                                  & \begin{tabular}[r]{@{}r@{}}3, ServerMem*: \\ 50MB/host,\end{tabular} & \begin{tabular}[r]{@{}r@{}}1, ClientMem*: 200MB/host, \\ RT OH*:10\%\end{tabular} & None                                                                & \begin{tabular}[r]{@{}r@{}}2, Alarms*: \\ 0.004/day/host\end{tabular} & None                                                           \\ 
    \hline
    E5                                                        & None                                                                 & None                                                                  & \begin{tabular}[r]{@{}r@{}}3, ServerMem*: \\ 30MB/host,\end{tabular} & \begin{tabular}[r]{@{}r@{}}1, ClientMem*: 100MB/host, \\ RT OH*:5\%\end{tabular}  & None                                                                & \begin{tabular}[r]{@{}r@{}}2, Alarms*: \\ 0.02/day/host\end{tabular}  & None                                                           \\ 
    \hline
    E6                                                        & \begin{tabular}[r]{@{}r@{}}5, Net: \\ 100MB/day/host\end{tabular} & \begin{tabular}[r]{@{}r@{}}6, Disk: \\ 15MB/day/host\end{tabular}  & \begin{tabular}[r]{@{}r@{}}3, ServerMem*: \\ 30MB/host,\end{tabular} & \begin{tabular}[r]{@{}r@{}}1, ClientMem*: 200MB/host, \\ RT OH*:1\%\end{tabular}  & \begin{tabular}[r]{@{}r@{}}4, Number of \\ nodes*: 100\end{tabular} & \begin{tabular}[r]{@{}r@{}}2, Alarms*: \\ 0.1/day/host\end{tabular}   & None                                                           \\ 
    \hline
    E7                                                        & \begin{tabular}[r]{@{}r@{}}5, Net: \\ 10MB/day/host\end{tabular}  & \begin{tabular}[r]{@{}r@{}}6, Disk: \\ 70MB/day/host\end{tabular}  & \begin{tabular}[r]{@{}r@{}}3, ServerMem*: \\ 20MB/host,\end{tabular} & \begin{tabular}[r]{@{}r@{}}1, ClientMem*: 50MB/host, \\ RT OH*:5\%\end{tabular}   & \begin{tabular}[r]{@{}r@{}}4, Number of \\ nodes*: 100\end{tabular} & \begin{tabular}[r]{@{}r@{}}2, Alarms*: \\ 0.1/day/host\end{tabular}   & None                                                           \\ 
    \hline
    E8                                                        & \begin{tabular}[r]{@{}r@{}}5, Net: \\ 42MB/day/host\end{tabular}  & \begin{tabular}[r]{@{}r@{}}4, Disk: \\ 100MB/day/host\end{tabular} & \begin{tabular}[r]{@{}r@{}}3, ServerMem*: \\ 26MB/host,\end{tabular} & \begin{tabular}[r]{@{}r@{}}2, ClientMem*: 250MB/host,\\ RT OH*:5\%\end{tabular}   & None                                                                & \begin{tabular}[r]{@{}r@{}}1, Alarms*: \\ 0.05/day/host\end{tabular}  & None                                                           \\ 
    \hline
    E9                                                        &\begin{tabular}[r]{@{}r@{}}4, Net: \\ 1MB/day/host\end{tabular}   & \begin{tabular}[r]{@{}r@{}}3, Disk: \\ 15MB/day/host\end{tabular}  & \begin{tabular}[r]{@{}r@{}}2, ServerMem*: \\ 10MB/host,\end{tabular} & \begin{tabular}[r]{@{}r@{}}1, ClientMem*: 150MB/host,\\ RT OH*:10\%\end{tabular}  & None                                                                & None                                                                     & None                                                           \\ 
    \hline
    E10                                                       & \begin{tabular}[r]{@{}r@{}}4, Net: \\ 100MB/day/host\end{tabular} & \begin{tabular}[r]{@{}r@{}}5, Disk:\\ 35MB/day/host\end{tabular}   & \begin{tabular}[r]{@{}r@{}}3, ServerMem*: \\ 30MB/host,\end{tabular} & \begin{tabular}[r]{@{}r@{}}1, ClientMem*: 100MB/host,\\ RT OH*:5\%\end{tabular}   & None                                                                & \begin{tabular}[r]{@{}r@{}}2, Alarms*: \\ 0.1/day/host\end{tabular}   & None                                                           \\ 
    \hline
    \begin{tabular}[r]{@{}r@{}}Reference\\ Range\end{tabular} & \begin{tabular}[r]{@{}r@{}}1$\sim$100MB\\ /day/host\end{tabular}     & \begin{tabular}[r]{@{}r@{}}15$\sim$100MB\\ day/host\end{tabular}      & 10$\sim$50MB/host                                                   & \begin{tabular}[r]{@{}r@{}}50$\sim$250MB/host, \\ 1$\sim$10\%\end{tabular}        & 10$\sim$100                                                         & \begin{tabular}[r]{@{}r@{}}0.001$\sim$0.1\\ /day/host\end{tabular}       & > 0.85                                                  \\
    \hline
    \end{tabular}}
    \end{table*}

%% file: tex/questionnaire.tex
\section{Online Questionnaire}
\label{sec:questionnaire}

\begin{table}[t!]
    \setlength{\abovecaptionskip}{5pt}
    \caption{Summarized results of online questionnaire}
    \begin{tabular}{|lr|}
    \hline
    \multicolumn{1}{|l|}{\textbf{Must-meet Factors}}       & \textbf{Summarized Result}                   \\ \hline
    \multicolumn{1}{|l|}{\textsf{Memory}}          &   < 20 MB/host           \\ \hline
    \multicolumn{1}{|l|}{\textsf{Client-side Overhead (RT OH)}}            &                     < 3 \%                 \\ \hline
    \multicolumn{1}{|l|}{\textsf{Client-side Overhead (ClientMem)}}            &                     < 100 MB/host                 \\ \hline 
    \multicolumn{1}{|l|}{\textsf{Interpretation}}            &  < 50 nodes   \\ \hline
    \multicolumn{1}{|l|}{\textsf{Alarm Triage}}            &            < 0.1 alarms/day/host                           \\ \hline
    \end{tabular}
    \label{tab:QualitativeResult}
\end{table}

\eat{
\begin{table}[t!]
    \setlength{\abovecaptionskip}{5pt}
    \caption{Summarized results of online questionnaire}
    \begin{tabular}{|lr|}
    \hline
    \multicolumn{1}{|l|}{\textbf{Key Factors}}       & \textbf{Summarized Result}                   \\ \hline
    \multicolumn{1}{|l|}{\textsf{Network}}      &  < 50 MB/day/host  \\ \hline
    \multicolumn{1}{|l|}{\textsf{Storage}}             &   < 75 MB/day/host           \\ \hline
    \multicolumn{1}{|l|}{\textsf{Memory}}          &   < 20 MB/host           \\ \hline
    \multicolumn{1}{|l|}{\textsf{Client-side Overhead (RT OH)}}            &                     < 3 \%                 \\ \hline
    \multicolumn{1}{|l|}{\textsf{Client-side Overhead (ClientMem)}}            &                     < 100 MB/host                 \\ \hline 
    \multicolumn{1}{|l|}{\textsf{Interpretation}}            &  < 50 nodes   \\ \hline
    \multicolumn{1}{|l|}{\textsf{Alarm Triage}}            &            < 0.1 alarms/day/host                           \\ \hline
    \multicolumn{1}{|l|}{\textsf{Accuracy}}            &         > 0.8        \\ \hline
    \end{tabular}
    \label{tab:QualitativeResult}
\end{table}
}

The interview participants helped us identify the key decision factors in adopting \ac{pedr} systems, and provided the reference ranges for the metrics used in these decision factors. 
However, these reference ranges were too coarse-grained and some participants did not provide their reference ranges for certain metrics. 
To eliminate research bias, we further designed an online questionnaire and recruited a broader scope of security engineers from a variety of companies to participate in our online questionnaire, and obtain more accurate reference values of these metrics.

\subsection{Participant Recruitment}
To recruit more professionals who have experiences with \ac{edr} systems, we disseminated our recruitment in two ways.
First, we asked the interviewees in Table~\ref{tab:Interview_backgroud} to help disseminate our recruitment information to their colleagues and security engineers who have experiences with \ac{pedr} and \ac{edr} systems. 
Second, we searched on business and employment-focused social media platforms such as LinkedIn~\cite{LinkedIn} and MaiMai~\cite{maimai} (the Chinese LinkedIn) to actively contact the people who are working on endpoint security. 
Specifically, we selected the keywords, ``Network Security'', ``Endpoint Detection and Response'', and ``Endpoint Security'', to narrow down the search scope to get the contacts of people that are knowledgeable with \ac{edr}. 
We phoned them first to introduce our purpose and know more about their backgrounds. 
For the qualified candidates, we invited them to participate in our online questionnaire if they were willing to get involved. 

\noindent\textbf{Participant Backgrounds}:
We invited 100+ participants in total, and 48 completed the questionnaires. 
Figure~\ref{fig:participants} in Appendix \S~\ref{appendix:questionnaire} shows the participants' backgrounds. 
The participants of our questionnaire come from companies in different industrial sectors, including government, IT technology, the security industry, financial services, manufacturing, etc. 
They all have experiences in enterprise security and have 4.4 years of the \ac{apt} combating experience on average.

\eat{
我们通过两种方式来招募受访者。
1.EDR tech leader 分发给组员。首先，我们通过各大公司的官网尝试联系到EDR的开发或者使用团队的技术经理。让技术经理分发给自己的组员。
2.求职平台。我们通过平台的筛选条件，网络安全，EDR，工程师，来缩小搜索范围到那些EDR的开发者和使用者。
为了增加大家填写的意愿，我们在招募书中写道，如果研究成功，PEDR系统的主要评价因素以及目前学术界和工业界的差距将会被识别出来。这些结果也许可以为他们提供PEDR开发和使用提供可能的好处。我们通过人工审核方式来剔除一些异常的回答。
}

\subsection{Questionnaire Survey Methodology}
We sent the online questionnaire link to the participants. 
Same as the participants of our interviews, we offered the participants of our online questionnaire to use our survey results for improving their uses of \ac{edr}/\ac{pedr} systems.
We controlled the length of the questionnaire within the range that the respondents can complete the questionnaire within 10 minutes to ensure a high response rate~\cite{SurveyDesign}. 
After we received the 48 responses, we further conducted attention-check to remove low-quality responses.
Specifically, we computed the average answering time and the average percentage of unknown answers in a response.
We then rejected a response if it was an outlier in terms of the answering time or the percentage of unknown answers.

\noindent\textbf{Questionnaire Questions:}
We design the questionnaire based on the results from the interview. 
In the interview, we have identified four must-meet factors: \textsf{Memory}, \textsf{Client-Side Overhead}, \textsf{Interpretation}, and \textsf{Triage}.
Thus, in the online questionnaire, we have four questions to determine the fine-grained reference values for these four must-meet factors, respectively.
For each question, we divide the reference range obtained in the interviews into five equal-sized sub-ranges and use five options to represent these sub-ranges.
In this way, we can obtain more fine-grained results for each decision factor.
Note that we also have an option ``I don't know''  to allow the participants to omit the questions that they are not confident about. 

We then followed the principles in \textit{How to Design and Frame a Questionnaire}~\cite{farooq2018design} to curate the questionnaire.
Specifically, we sent our questionnaire to the interview participants and asked them to confirm that our question design follows their interview answers and that descriptions can be easily understood. 
Appendix \S~\ref{appendix:online} presents the detailed questionnaire.

\noindent\textbf{Data Processing:} 
We use a SaaS-based questionnaire platform for managing the questionnaire data. 
Two authors independently computed questionnaire data statistics and cross-checked the results. 
The questionnaire data retention is one year, same as that of the interview data.




\begin{figure}[h]
    \setlength{\abovecaptionskip}{5pt}
    \includegraphics[width=0.49\textwidth]{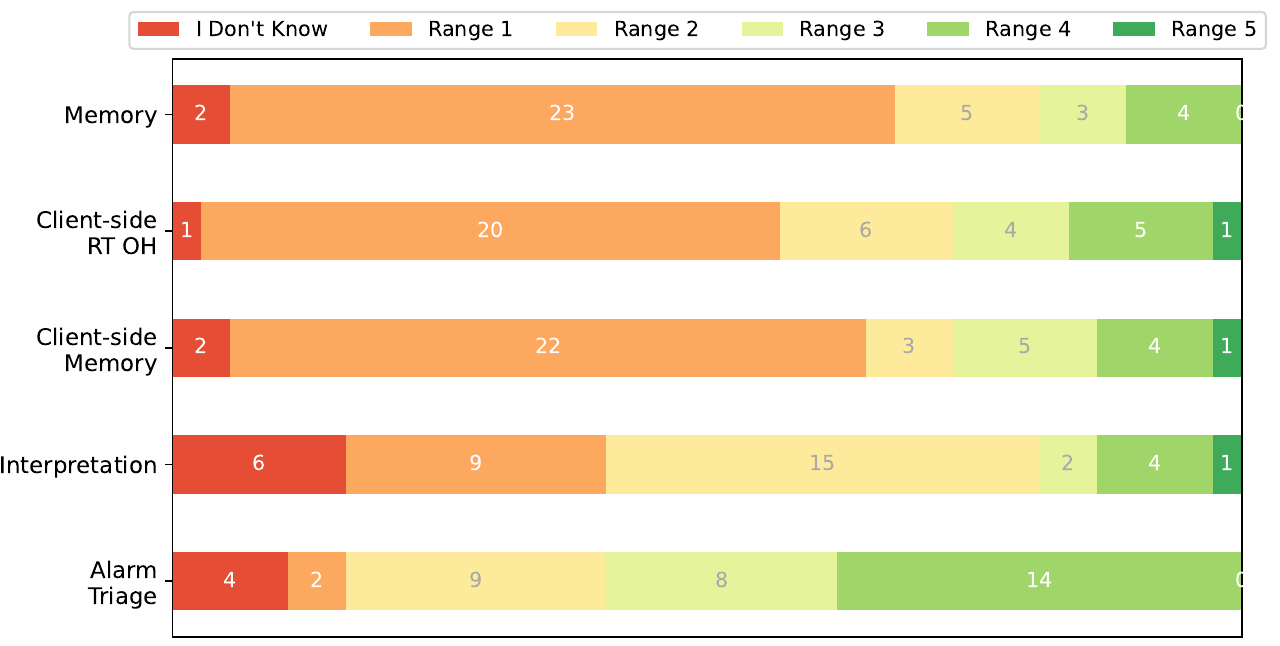}
    \caption{Metric Results of Our Questionnaire Study.}
    \label{fig:questionnaire-results}
\end{figure}

\subsection{Results}
In total, we received 48 questionnaire responses. 
The average answering time was 7 minutes) and the average percentage of unknown answers was 14\%.
We rejected 11 responses as they were outliers in terms of the answering time (<100 seconds) or the percentage of unknown answers (>50\% of the answers). 
Thus, we had 37 valid responses, and the distribution of the answers for each option is shown in Figure~\ref{fig:questionnaire-results}.
For each key factor, we chose the option selected by the largest number of participants as the reference value. 
We summarize the results of the reference values in Table~\ref{tab:QualitativeResult}.
These results are later used to guide our literature survey and measurement study.

%% file: tex/literature.tex
\section{Literature Survey}
\label{sec:LiteratureSurvey}
The research objective of our literature survey is to investigate the \ac{pedr} systems described in recent publications and check whether they can satisfy the four must-meet factors (\textsf{Client-Side Overhead}, \textsf{Memory}, \textsf{Interpretation}, and \textsf{Alert Triage}). 
We try to report the exact values in the original papers for each decision factor. 

\subsection{Methodology}
We systematically inspected all the provenance analysis papers published in top conferences and journals during 2017-2022, including IEEE S\&P, USENIX Security, CCS, NDSS, ACSAC, TDSC, and TIFS. 
We carefully read the abstracts of these papers to classify whether they are in our scope. 
Finally, we selected 20 papers on \ac{pedr} systems and classified their approaches into rule-based approaches (5), anomaly-based approaches (7), and investigation approaches (8).

For these 20 approaches, we investigate whether they have been evaluated against the seven decision factors.
Specifically, two authors independently inspected the evaluation results of these 20 papers on the four must-meet factors and cross-checked the results.
Table~\ref{tab:literature} shows the reported values for each paper.
For the systems that were evaluated on different datasets, we calculate the average values on different datasets listed in their papers by default. 
For \textsf{Accuracy}, we cannot simply use their average reported values, and thus we summarized the range of the reported values. 
We use ``$-$'' to indicate an approach was not evaluated against a decision factor. 

\subsection{Surveyed Papers}
We can roughly divide existing approaches into two categories: \textit{Rule-based Detection} and \textit{Anomaly-based Detection}~\cite{SoK-History}. 

\begin{itemize}[noitemsep, topsep=1pt, partopsep=1pt, listparindent=\parindent, leftmargin=*]
\item Rule-based detection approaches leverage prior expert knowledge and experience of attacks to design policies for event matching and behavior extraction. 
Tag propagation and rule matching are the two most commonly used methods. 
SLEUTH~\cite{hossain2017sleuth} is the first provenance-based tag policy framework that assigns trustworthiness and confidentiality tags to system entities and propagates on the provenance graph. MORSE~\cite{hossain2020morse} is designed based on SLEUTH with refined policies to reduce the amount of false positive alarms. HOLMES~\cite{milajerdi2019holmes} and RapSheet~\cite{hassan2020rapsheetl} leverage the MITRE ATT\&CK knowledge-base to configure their rules, mapping low-level events to high-level \ac{ttp}, \ac{hsg} and \ac{tpg} for attack detection and investigation. Pagoda~\cite{8450016} combines the abnormality of a single path and the entire provenance graph.

\item Anomaly-based detection systems are diverse in strategies. Overall, they always learn normal behaviors from historical data and treat deviations from normal behavior as malicious. Both StreamSpot~\cite{DBLP:journals/corr/ManzoorMVA16} and UNICORN~\cite{han2020unicorn} extract the provenance graph into sketches, a vector, as features for clustering and label the outliers as anomalies. UNICORN chooses StreamSpot as the baseline and uses the public dataset collected by StreamSpot, achieving better performance. ProvDetector~\cite{wang2020provdetector} leverages probability density-based Local Outlier Factors to detect stealth malware paths, embedded into fixed length vectors using graph embedding methods.  ZePro~\cite{8327913} uses Bayesian Networks for zero-day attack path identification, and P-Gaussian~\cite{8935406} uses Gaussian Distribution for sequence similarity detection. Poirot needs to manually design \ac{qg}, generated from \ac{cti} report with pre-known expert knowledge. SHADEWATCHER~\cite{9833669} extracts the interaction from the provenance graph and constructs a recommendation model for learning to classify system entity interactions into normal and adversarial.
\end{itemize}

\input{tab/literature.tex}

\subsection{Results}
We next report our analysis of these 20 approaches for each of the must-meet factors.

\noindent\textbf{\textsf{Client-Side Overhead}}.
We found that only one paper (RTAG) provided evaluations against the client-side overhead.
RTAG is an improvement on RAIN, that implemented the system logging logic with comprehensive semantics to record whole-system activities to enable cross-host attack investigation. 
It mainly measured the runtime overhead and compared it with existing full-system provenance systems. 
The other 19 systems focus on building detection and investigation algorithms and rely on third-party collectors to monitor provenance data. 
Thus, \textit{these papers omit the evaluations of the client-side overhead introduced by the collectors}.

\begin{table}[t]
    \setlength{\abovecaptionskip}{5pt}
    \centering
    \caption{List of existing provenance data collectors. ``RT OH'' is the average runtime overhead of applications in their evaluations. ``Mem'' is the average memory consumption.}
    \label{tab:collectors}
    \resizebox{0.49\textwidth}{!}{\begin{tabular}{|l|l|p{80pt}|p{40pt}|r|r|} 
    \hline
                   & \textbf{Platform} & \textbf{Owner}                    & \textbf{Affect}  & \textbf{RT OH (\%)} & \textbf{Mem (MB)}  \\ \hline
    Sysdig~\cite{sysdig}         & Linux    & Sysdig.Inc               &    \cite{depcomm,depimpact}     &   NA      &   NA   \\ \hline
     \multirow{3}{*}{Auditd~\cite{auditd}}   & \multirow{3}{*}{Linux}    & \multirow{3}{*}{Linux Foundation}         &     \cite{hossain2017sleuth,hossain2020morse,milajerdi2019holmes,milajerdi2019poirot,zengy2022shadewatcher,Kwon2018MCIM,liu2018PRIOTRACKER,hassan2019nodoze,10.1145/3564625.3567997}   &   \multirow{3}{*}{NA}      &   \multirow{3}{*}{NA}   \\ \hline
     \multirow{2}{*}{DTrace~\cite{10.5555/1247415.1247417,wang2019dtrace}} &    \multirow{2}{*}{Linux}        &  \multirow{2}{*}{Sun Microsystems}  &    \cite{hossain2020morse,milajerdi2019holmes,milajerdi2019poirot,Kwon2018MCIM,10.1145/3564625.3567997}  &  \multirow{2}{*}{3.2}      &  \multirow{2}{*}{NA} \\ \hline
     \multirow{2}{*}{Camflow~\cite{camflow}}        & \multirow{2}{*}{Linux}    & University of Cambridge  &   \multirow{2}{*}{\cite{han2020unicorn}}      &   \multirow{2}{*}{9.7}       &   \multirow{2}{*}{NA}   \\ \hline
    LTTng~\cite{desnoyers2008lttng}          & Linux    & EfficiOS                 &   NA     &     NA    &   NA   \\ \hline
    \multirow{2}{*}{ETW~\cite{etw}}            & \multirow{2}{*}{Windows} & \multirow{2}{*}{Microsoft}                &   \cite{milajerdi2019holmes,milajerdi2019poirot,Kwon2018MCIM,liu2018PRIOTRACKER,hassan2019nodoze,10.1145/3564625.3567997}  &  \multirow{2}{*}{NA}       &   \multirow{2}{*}{NA}   \\ \hline
    KennyLoggings~\cite{logging2020ccs} & Linux    & UIUC                     &   NA       &   4.6      &    NA  \\ \hline
    Hardlog~\cite{ahmad2022hardlog}        & Linux    & Microsoft                &   NA       &   6.3      &   NA   \\ \hline
    \multirow{2}{*}{Quicklog~\cite{281386}}       & \multirow{2}{*}{Linux}    & Florida State University &   \multirow{2}{*}{NA}        &  \multirow{2}{*}{5.3}       &   \multirow{2}{*}{NA} \\ \hline
    SystemTap~\cite{jacob2008systemtap,eigler2005architecture} &     Linux       &        Linux Foundation            &    \cite{DBLP:journals/corr/ManzoorMVA16}         &   NA    &  NA \\ \hline
    \multirow{2}{*}{RAIN~\cite{10.1145/3133956.3134045}} &      \multirow{2}{*}{Linux}  & Georgia Institute of Technology & \multirow{2}{*}{\cite{10.1145/3133956.3134045,Ji2018EnablingRC}} & \multirow{2}{*}{NA} & \multirow{2}{*}{NA} \\ \hline
    Karma~\cite{inproceedings,simmhan2006performance} &  Linux  &  Indiana University   &  \cite{8450016}   & NA   &NA \\ \hline
    PASS~\cite{10.5555/1267359.1267363} & Linux   &  Harvard University  &  \cite{8450016}  &  10.5    &  NA \\ \hline
    \end{tabular}}
    \end{table}

Due to the lack of evaluations on provenance collectors in these papers, we further surveyed the available provenance collectors used in industry and academia and listed them in Table~\ref{tab:collectors}. 
Unfortunately, we found that there were no systematic evaluations of the client-overhead introduced by existing provenance collectors.
Although there were six collectors that had evaluated the runtime overhead, the three most commonly used collectors, Sysdig, Auditd, and ETW, did not have evaluations of their introduced overheads on other client-side applications. 
Even worse, none of the existing provenance collectors can satisfy the reference value of runtime overhead (< 3\%). 
Moreover, we found no evaluations of the memory consumption for these collectors. 
Therefore, we further carried out a measurement study on these three collectors in Section~\ref{sec:measurementclientside}.

\noindent\textbf{\textsf{Memory}}.
8 out of 20 approaches have evaluated the memory consumption on the server side, but none of them directly report the average memory consumption for each monitored machine, and we calculate this value by dividing the overall memory consumption by the number of hosts in their datasets.
The results show that most of the reported values are much higher than the reference values we obtained (< 20MB/host), except for RAPID~\cite{10.1145/3564625.3567997}. 
Particularly, SHADEWATCHER exceeds the expected value by 209 times, which indicates that it can hardly be deployed in the industrial environment.
Although RAPID can satisfy the requirement, it needs to utilize third part detection systems for investigation. 
These results indicate that there is a gap in the methodology of memory consumption evaluation between academia and industry.

\noindent\textbf{\textsf{Interpretation}}.
The investigation cost is measured by the size of generated provenance graphs.
This metric is well-evaluated by existing systems.
However, the results vary from 12 nodes to $1.76\times 10^5$ nodes as this factor is highly correlated with the system design.
In general, rule-based systems, such as SLEUTH, MORSE, and HOLMES, can generate smaller provenance graphs in alarms than anomaly-based systems, such as StreamSpot and UNICORN. 
Nevertheless, most of the rule-based and anomaly-based systems have to optimize their interpretation cost by 1 - 4 orders of magnitude in order to meet the industrial requirements ($<$50 nodes). 

\noindent\textbf{\textsf{Alarm Triage}}.
None of the papers provide evaluations for the cost of alarm triage.
In fact, all these papers ignore the factor of triage cost when they evaluate the accuracy of a \ac{pedr} system. Note that in practice, the ratio of attack-related data is very low (less than 0.1\%)~\cite{hassan2019nodoze,liu2018towards}. 
Thus, even though they can achieve high accuracy (0.95 averaged from their best-reported values), the triage costs are usually not acceptable in practice. 

\subsection{Summary}
Our literature survey shows that almost all the existing systems do not provide evaluations against the four must-meet factors.
Interpretation cost is the only factor that is evaluated by most of the papers, while a large proportion of the surveyed systems cannot meet the requirement from the industry.
Similarly, a small set of papers provide evaluations for part of the four factors, and their results show that these systems fail to satisfy the reference values obtained from our studies.

%% file: tab/literature.tex
\begin{table*}[h]
    \setlength{\abovecaptionskip}{5pt}
    \centering
    \caption{Summarization of literature review. The values are directly reported or averaged the values on different datasets listed in their paper. The empty cells mean the corresponding paper does not evaluate the corresponding decision factor.}
    \label{tab:literature}
     \resizebox{0.99\textwidth}{!}{\begin{tabular}{|l|l|r|r|r|r|r|r|r|r|r|r|} 
    \hline
    \multirow{3}{*}{\textbf{Type}}& \multirow{3}{*}{\textbf{Tool Name}} &               \multicolumn{3}{c|}{\textbf{Client-side Overhead}}                                                                                                            & \multirow{3}{*}{\begin{tabular}[c]{@{}c@{}}\textbf{Storage}\\\textbf{(/MB/host/day)}\end{tabular}} & \multirow{3}{*}{\begin{tabular}[c]{@{}c@{}}\textbf{Memory}\\\textbf{(MB/host)}\end{tabular}} & \multirow{3}{*}{\begin{tabular}[c]{@{}c@{}}\textbf{Alarm Triage}\\\textbf{(\#Alarm/host/day)}\end{tabular}} & \multicolumn{1}{c|}{\multirow{3}{*}{\begin{tabular}[c]{@{}c@{}}\textbf{Interpretation}\\\textbf{(\#Node, \#Edge) }\end{tabular}}} & \multirow{3}{*}{\textbf{Precision}} & \multicolumn{1}{c|}{\multirow{3}{*}{\textbf{Recall}}} & \multirow{3}{*}{\textbf{Accuracy }}  \\ 
    \cline{3-5}
                                        & & \multicolumn{1}{c|}{\textbf{Agent}} & \textbf{RT OH(\%)} & \begin{tabular}[c]{@{}c@{}}\textbf{ClientMem}\\\textbf{(MB)}\end{tabular} &                                                                                                    &                                                                                                &                                                      &                                                                                                              &                                      &                                   &                                      \\ 
    \hline
    \multirow{12}{*}{\textbf{Detection}}& \multicolumn{1}{l|}{SLEUTH~\cite{hossain2017sleuth}}        & Auditd                                                       & -                &    -                                                                    & 362.87                                                                                             & 81.93                                                                                             &         -           & (52, -)                                                                                                      &                       -               &-                                   & -                                     \\ 
    \cline{2-12}
    &MORSE~\cite{hossain2020morse}                               & Auditd, DTrace                                                                  & -                &   -                                                                     & 1266.67                                                                                            & 230.4                                                                                           &       -           & (283, -)                                                                                                     &                       $\approx 0$  &     1.00                               &             -                         \\ 
    \cline{2-12}
    &HOLEMS~\cite{milajerdi2019holmes}                              & Auditd, Dtrace, ETW                                                              & -                &   -                                                                     & 179.23                                                                                             & 104.76                                                                                         & -                                                    & (-, 400)                                                                                                     &                      1.00                &1.00                                   &1.00                                      \\ 
    \cline{2-12}
    &RapSheet~\cite{hassan2020rapsheetl}                            & Symantec \ac{edr}                                      & -                &     -                                                                   & 358.00                                                                                             & -                                                                                              & -                                                & (12, 39)                                                                                                     &                       0.26               &1.00                                   &0.75 - 0.95                                      \\ 
    \cline{2-12}
    &Pagoda~\cite{8450016}                              & Karma~\cite{inproceedings}, PASS~\cite{10.5555/1267359.1267363}                                        & -                &         -                                                               & 1126.40                                                                                            & -                                                                                           & -                                                    & (13315, 10964)                                                                                               &                          0.92-1.00            &1.00                                   &    0.75 - 0.95                                  \\ 
    \cline{2-12}
    &StreamSpot~\cite{DBLP:journals/corr/ManzoorMVA16}                          & SystemTap~\cite{jacob2008systemtap}                                                                & -                &      -                                                                  & -                                                                                                  & -                                                                                           & -                                                    & (8315,173857)                                                                                                &                        0.50-1.00              & -                                  &  0.50 - 0.80                                    \\ 
    \cline{2-12}
    &UNICORN~\cite{han2020unicorn}                             & CamFlow~\cite{10.1145/3127479.3129249}                                                                  & -             &   -                                                                     & 24917.33                                                                                           & -                                                                                              & -                                                    & $(1.76\times 10^5,2.82\times 10^6)$                                       & 0.80 - 0.99                                     &           0.88 - 1.00                        &             0.84 - 0.99                         \\ 
    \cline{2-12}
    &ProvDetector~\cite{wang2020provdetector}                        & -                                                                 & -                &    -                                                                    & -                                                                                                  & -                                                                                              & -                                                    & (-, -)                                                                                                       &                      0.96                &              0.99                     &        -                              \\ 
    \cline{2-12}
    &ZePro~\cite{8327913}                               & -                                                                  & -                &    -                                                                    & 266.67                                                                                             & 57.14                                                                                          & -                                                    & (1853, 2249)                                                                                                 &                         -             &     -                              &    -                                  \\ 
    \cline{2-12}
    &P-Gaussian~\cite{8935406}                          & -                                                                 & -               &    -                                                                    & 864                                                                                                & 152.5                                                                                          & -                                                    & (1949, 3045)                                                                                                 &                         -             &     0.66 - 0.94                              &      0.65 - 0.95                                \\ 
    \cline{2-12}
    &Poirot~\cite{milajerdi2019poirot}                              & Auditd, Dtrace, ETW                                                                 & -                &   -                                                                     & 6500.55                                                                                            & 122.39                                                                                         & -                                                    & (-, -)                                                                                                       &                      1.00                &            1.00                       & 1.00                                      \\ 
    \cline{2-12}
    &SHADEWATCHER~\cite{zengy2022shadewatcher}                        & Auditd                                                                  & -               &      -                                                                  & 59112.73                                                                                           & 4194.30                                                                                           & -                   & (-, -)                                                                                                       &                      0.86 - 1.00                & 0.95 - 1.00                                   & 0.98 - 1.00                                      \\
    \hline
    \multirow{8}{*}{\textbf{Investigation}}& RTAG~\cite{Ji2018EnablingRC}     & RAIN                           &   4.84               &    -                                                                    & 1536 - 4096                                                                                            & -                                                                                             & -                  & (164.67, 3200)                                                                                                      &                 -              &-                                   & 1.00                                     \\ 
    \cline{2-12}
    & MCI~\cite{Kwon2018MCIM}     & Auditd, Dtrace, ETW                           &   -               &    -                                                                    & -                                                                                            & -                                                                                             & -                  & (34.56, 62.87)                                                                                                      &                 0.92- 1.00              & 0.95 - 1.00                                  & -                                     \\ 
    \cline{2-12}
    &  PrioTracker~\cite{liu2018PRIOTRACKER}     & Auditd, ETW                                                        & -                &    -                                                                    & 998.64                                                                                             & -                                                                                             & -                  & (-, 1219)                                                                                                      &                       -               &-                                   & -                                     \\ 
    \cline{2-12}
    & NoDoze~\cite{hassan2019nodoze}                             & Auditd, ETW                                                                  & -                &   -                                                                     & 428.90                                                                                            & -                                                                                           & -                  & (14, 14)                                                                                                     &   0.50                  &     1.00                               &             0.86                         \\ 
    \cline{2-12}
    & ATLAS~\cite{263852}                             & -                                                              & -                &   -                                                                     & 2286.93                                                                                             & -                                                                                         & -                                                    & (-, -)                                                                                                     &                      0.91                &0.97                                   &0.99                                      \\ 
    \cline{2-12}
    & DEPCOMM~\cite{depcomm}                            & Sysdig                                                              & -                &   -                                                                     & -                                                                                             & -                                                                                         & -                                                    & (289, -)                                                                                                     &                      -                &-                                   &-                                      \\ 
    \cline{2-12}
    & DEPIMPACT~\cite{depimpact}                             & Sysdig                                                              & -                &   -                                                                     & -                                                                                             & -                                                                                         & -                                                    & (-, 234.27)                                                                                                     &                      0.79 - 0.85                & 1.00                                   &-                                      \\ 
    \cline{2-12}
    & RAPID~\cite{10.1145/3564625.3567997}                             & Auditd, Dtrace, ETW                                                              & -                &   -                                                             & 4743.40                                                                                             & 30.04                                                                                         & -                                                    & (-, -)                                                                                                     &                -                & -                                  &-                                      \\ 
    \hline
    
    \end{tabular}}
    \end{table*}

%% file: tex/evaluation.tex
\section{Measurement Study}
\label{sec:EmpiricalStudy}
To better understand whether existing \ac{pedr} systems can satisfy the four must-meet factors and how much improvement on these factors is needed, we conducted a measurement study on the representative systems described in the surveyed papers and obtained their metric values to compare with the collected reference values.
We next describe our measurement study on both the client-side and the server-side.


\subsection{Client-Side Measurement Study}
\label{sec:measurementclientside}
In this section, we empirically study the \textsf{Client-Side Overhead} factor using three representative collectors.
We deployed these collectors to hosts with different hardware configurations
and also measured their introduced overheads on seven representative applications used in our surveyed papers.

\noindent\textbf{Representative Collectors.}
In our measurement study, we chose three most widely used industrial open-source collectors, Sysdig, LTTng, and Auditd, from the collectors listed in Table~\ref{tab:collectors}.
These three collectors are adopted by the majority of the existing \ac{pedr} systems.
They have industrial quality and are actively maintained. 
We excluded DTrace because it has a similar performance to Sysdig, and it requires significant technical knowledge to utilize and optimize, which may cause potential bias~\cite{sysdigvsdtrace}. 
We did not measure the client-side overhead of ETW due to two reasons. First, we found no way to turn off the kernel module of ETW completely. 
Second, ETW does not have an official user-space collector, and our study of it could be significantly biased. We also excluded other collectors because they are outdated and lack downstream users.  

\noindent\textbf{Representative Applications.}
We chose seven representative applications used in the surveyed papers, which can be classified into two categories: 
\begin{itemize}[noitemsep, topsep=1pt, partopsep=1pt, listparindent=\parindent, leftmargin=*]
\item \textit{I/O-intensive applications}: We first chose commonly used applications of C++, including Nginx~\cite{reese2008nginx}, Redis~\cite{redis}, Postmark~\cite{katcher1997postmark}. 
We also chose two applications of other languages, namely Django~\cite{py-django} for Python and http~\cite{go-http} for Golang. 
\item \textit{CPU-intensive applications}: We chose OpenSSL~\cite{openssl} and 7-ZIP~\cite{7zip}. 
\end{itemize}


\noindent\textbf{Experiment Setup.}
In our measurement study, we followed the minimal workload principle to avoid possible biases introduced by extra provenance data processing. 
We simply directed the three collectors to dump their collected data into a file in an in-memory file system. 
Note that this protocol measures the lower bound of the client-side overhead of the provenance collectors as they usually contain more complex processing logic or need to dump data into much slower devices, such as networks and hard disks.
Therefore, we expect the real client-side overhead of \ac{pedr} systems should be higher than the values we reported in this paper. 
We ran experiments on these three tools under four hardware configurations with different numbers of cores and different sizes of memory on both virtual and physical machines, as shown in Table~\ref{tab:harwareconfig}.

We ran these applications using their official benchmarks while measuring their performance. 
Specifically, we used \textit{wrk}~\cite{wrk} with 1,000 concurrent connections to benchmark Nginx. For Redis, we used the \textit{redis-benchmark} configured to send 1,000,000 requests and measure the speed of operation \code{get}. We used the built-in benchmark with the configuration of manipulating 500 files concurrently and launching 100,000 transactions to evaluate the performance of PostMark. We used the Phoronix Test Suite, one of the most comprehensive
benchmark suites of web applications~\cite{phoronix,sharath-ijca-13}, to benchmark Django and http. For OpenSSL, we relied on the default \textit{speed} benchmark configured to utilize all CPU cores and measure the time to compute one rsa4096 signature. For 7-ZIP, we used the built-in benchmark configured to utilize all CPU cores and measure the compression speed in MIPS. We repeated each experiment ten times and reported the average metric values of the benchmarks.

\noindent\textbf{\textsf{Runtime Overhead.}}
 We show the experiment results in Table~\ref{tab:benchmarkidle}. We notice that for I/O-intensive applications, there are relatively high overheads compared to the case without turning on the collectors. We also notice that as the number of CPU cores increases, the overhead decreases. This is because all the collectors are single-threaded, which can only utilize one CPU core. When the number of CPU cores is small, the collectors will compete for the resources with the applications. 
 For example, for the single-core machines (C1 and C5), the collectors add at most 821\% more overhead to Nginx.  
 Particularly, we find that Auditd introduces a significant overhead because it uses Netlink and has heavy processing logic. For CPU-intensive applications, the overheads compared to the case without turning on the collectors are much smaller, which is less than 3\% on average across all configurations. To conclude, all the provenance collectors introduce inevitable overhead compared to the case without turning on the collectors since they record and consume the provenance events. 

\noindent\textbf{\textsf{Memory.}}
The memory consumption of the collectors is listed in Table~\ref{tab:collectormem}.
Memory consumption of collectors consists of two parts: user-mode and kernel-mode memory. For Sysdig, LTTng, and Auditd, the user-mode memory consumption is 30M, 15.9M, and 1.9M, respectively, which is independent of hardware configurations and applications. 
For kernel-mode memory cost, Auditd allocated a fixed size buffer of 64MB by default; 
Sysdig and LTTng allocated a fixed size buffer for each core, 8MB and 2MB respectively.

\begin{table}[]
\setlength{\abovecaptionskip}{5pt}
\caption{Hardware configurations for our measurement study}
\label{tab:harwareconfig}
\resizebox{0.48\textwidth}{!}{\begin{tabular}{|c|r|r|r|r|}
\hline
\multirow{2}{*}{\begin{tabular}[c]{@{}c@{}}\textbf{Physical }\\\textbf{Machine}\end{tabular}} & \multicolumn{1}{c|}{\textbf{C1}}   & \multicolumn{1}{c|}{\textbf{C2}}  & \multicolumn{1}{c|}{\textbf{C3}}   &\multicolumn{1}{c|}{\textbf{C4}}   \\ \cline{2-5}
 &1CPU + 2GB & 4CPU + 8GB& 16 CPU + 32GB& 32 CPU + 64GB\\ \hline
 \multirow{2}{*}{\begin{tabular}[c]{@{}c@{}}\textbf{Virtual }\\\textbf{Machine}\end{tabular}}&\multicolumn{1}{c|}{\textbf{C5}}   & \multicolumn{1}{c|}{\textbf{C6}}   & \multicolumn{1}{c|}{\textbf{C7}}   & \multicolumn{1}{c|}{\textbf{C8}}   \\ \cline{2-5}
 &1CPU + 2GB& 4CPU + 8GB& 16 CPU + 32GB& 32 CPU + 64GB\\ \hline

\end{tabular}}
\end{table}

\begin{table*}
    \setlength{\abovecaptionskip}{5pt}
    \centering
    \caption{\textbf{Application benchmarking results. We measure the processing time per request/transaction seven representative applications. We report the median values across 10 runs. All values are shown as the relative runtime overhead (\%).}}
    \label{tab:benchmarkidle}
    \begin{tabular}{|l|l|r|r|r|r|r|r|r|r|r|} 
    \hline
        \textbf{Application}                 & \textbf{Collector}  & \textbf{C1} & \textbf{C2} & \textbf{C3} & \textbf{C4} & \textbf{C5}      & \textbf{C6}      & \textbf{C7}       & \textbf{C8}     & \textbf{Avg}     \\ 
    \hline
    \multirow{3}{*}{Nginx}           & Auditd  & 597.30 & 101.30 & 34.60 & 34.80 & 821.10 & 186.30 & 23.70 & 10.90  & 226.25 \\ 
    \cline{2-11}
                                     & Sysdig  & 70.20 & 26.10 & 14.60 & 15.60 & 68.10 & 21.20 & 9.50 & 7.20   & 29.06 \\ 
    \cline{2-11}
                                     & LTTng & 24.80 & 10.70 & 10.00 & 11.70 & 26.30 & 25.80 & 7.00 & 1.40   & 14.71 \\ 
    \hline
    \multirow{3}{*}{Redis}           & Auditd  & 457.00& 58.10 & 41.70 & 50.20& 512.00 & 53.20 & 46.00 & 43.20 &157.67\\ 
    \cline{2-11}
                                     & Sysdig & 17.90 & 20.00 & 17.20 & 16.20 & 21.00 & 16.40 & 15.60 & 5.70   &  16.25   \\ 
    \cline{2-11}
                                     & LTTng  & 8.30 & 8.40 & 10.00 & 5.10 & 13.60 & 6.90 & 1.40 & 2.70 &  7.05   \\ 
    \hline
    \multirow{3}{*}{Postmark}        & Auditd   & 406.00 & 81.80 & 84.30 & 78.40 & 658.00 & 149.40 & 157.20& 116.20 &216.41 \\ 
    \cline{2-11}
                                     & Sysdig  & 88.80 & 19.20 & 18.00 & 22.00 & 98.80 & 23.20 & 16.50 & 7.50        & 36.75\\ 
    \cline{2-11}
                                     & LTTng   & 10.30 & 9.40 & 12.30 & 18.10 & 12.90 & 10.30 & 10.90 & 11.60     & 11.98 \\ 
    \hline
    \multirow{3}{*}{Django (Python)} & Auditd  & 2.50 & 0.70 & 2.10 & 2.30 & 1.20 & 0.50 & 1.50 & 2.10   &1.62\\ 
    \cline{2-11}
                                     & Sysdig   & 1.00 & 1.00 & 0.40 & 1.10& 1.10 & 1.40 & 0.10 & 0.30     & 0.80 \\ 
    \cline{2-11}
                                     & LTTng  & 1.70 & 2.10 & 1.70& 1.00 & 1.20 & 0.30 & 0.80 & 1.10      &1.24\\ 
    \hline
    \multirow{3}{*}{http (Golang)}   & Auditd   & 341.00 & 97.30 & 31.20 & 11.30 & 516.00 & 91.60& 35.30 & 15.50 &142.40 \\ 
    \cline{2-11}
                                     & Sysdig & 60.70 & 13.90 & 10.60 & 2.80& 76.70& 11.90 & 4.10 & 2.20    & 22.86 \\ 
    \cline{2-11}
                                     & LTTng  & 13.80 & 6.50 & 4.20 & 4.10 & 13.40 & 6.20 & 5.80 & 4.20        & 7.28 \\ 
    \hline
    \multirow{3}{*}{OpenSSL}         & Auditd  & 2.90 & 1.80 & 1.20 & 1.00 & 6.90 & 0.10 & 1.70 & 0.20 &1.98\\ 
    \cline{2-11}
                                     & Sysdig & 0.50 & 0.80 & 0.40 & 0.10 & 0.50 & 1.40 & 0.30& 0.10 & 0.51     \\ 
    \cline{2-11}
                                     & LTTng  & 2.50 & 0.50 & 0.10 & 0.10& 0.20 & 0.20 & 1.70 & 0.60  &  0.74   \\ 
    \hline
    \multirow{3}{*}{7-ZIP}           & Auditd  & 17.40 & 10.90 & 5.40 & 3.70 & 16.90 & 5.60 & 2.40 & 2.00  &8.04\\ 
    \cline{2-11}
                                     & Sysdig & 1.50 & 1.30 & 1.10 & 1.10 & 1.20 & 1.00 & 0.80 & 0.70   & 1.08   \\ 
    \cline{2-11}
                                     & LTTng & 2.40 & 1.80 & 0.90 & 0.80 & 4.70 & 2.30 & 0.10 & 0.10  &  1.64 \\
    \hline
    \end{tabular}
    \end{table*}

\begin{table}[t]
\setlength{\abovecaptionskip}{5pt}

 \caption{Memory consumption of provenance data collectors}
    \label{tab:collectormem}
\begin{tabular}{|l|r|r|r|r|}
\hline
\textbf{Agent}  & \textbf{C1/C5}   & \textbf{C2/C6}   & \textbf{C3/C7}   & \textbf{C4/C8}   \\ \hline
Auditd  & 65.9M & 65.9M & 65.9M & 65.9M \\ \hline
Sysdig & 38M   & 62M   & 158M  & 286M  \\ \hline
LTTng  & 17.9M & 23.9M & 47.9M & 79.9M \\ \hline
\end{tabular}
\end{table}




\subsection{Server-Side Measurement Study}
\label{sec:measurementserverside}
In this section, we empirically study the \textsf{Memory}, \textsf{Interpretation}, and \textsf{Triage} factors 
using three representative \ac{edr} systems.

\noindent\textbf{Representative \ac{pedr}s}.
We chose ProvDetector~\cite{wang2020provdetector}, UNICORN~\cite{han2020unicorn}, and HOLMES~\cite{milajerdi2019holmes} due to the following reasons.
First, these three systems have the highest precision according to Table~\ref{tab:literature}. 
Since the average number of alarms approximates the precision, we expected these three systems to have the lowest number of alarms per host per day. 
Second, these three systems cover the two categories of \ac{pedr} systems. HOLMES is one of the state-of-the-art rule-based systems, while UNICORN and ProvDetector are the two leading learning-based systems. We cannot implement MORSE~\cite{9152772}, Poirot~\cite{poirot}, and RapSheet~\cite{hassan2020rapsheetl} because they rely on unpublished rules or CTI reports.
We fail to implement SHADEWATCHER~\cite{zengy2022shadewatcher} because it depends on an unpublished recommendation model. We omit  SLEUTH~\cite{hossain2017sleuth} and StreamSpot~\cite{DBLP:journals/corr/ManzoorMVA16} because they are inferior to HOLMES and UNICRON, respectively. 
We exclude pagoda, ZePro, and P-gaussian because they adopt similar techniques as ProvDetecor, and ProvDetecor is the most recognized approach among them. We also exclude the investigation systems like PrioTracker~\cite{liu2018PRIOTRACKER}, NoDoze~\cite{hassan2019nodoze}, ATLAS~\cite{263852}, and DEPCOMM~\cite{depcomm} because they need to work with unpublished third-party attack detection tools.  
We implemented HOLMES using the detection rules provided in its paper and configured it to achieve the best performance based on our empirical knowledge. 
We implemented ProvDetector according to the description in its paper and adopted its default configurations. 
We directly used the published source code of UNICORN and adopted its default parameters but modified its data parser to accept our data format.
We conducted our experiments on the following datasets. 

\begin{table}[t!]
    \setlength{\abovecaptionskip}{5pt}
    \centering
    \caption{Overview of the evaluation datasets}
    \resizebox{0.48\textwidth}{!}{
    \begin{tabular}{|l|r|r|r|r|r|r|}
    \hline
    \multicolumn{1}{|c|}{\multirow{2}{*}{\textbf{Dataset}}} & \multicolumn{1}{c|}{\multirow{2}{*}{\begin{tabular}[c]{@{}c@{}}\textbf{Host}\\\textbf{Num}\end{tabular}}} &\multirow{2}{*}{\textbf{Days}}& \multicolumn{1}{c|}{\multirow{2}{*}{\begin{tabular}[c]{@{}c@{}}\textbf{Data}\\\textbf{Size}\end{tabular}}} & \multicolumn{1}{c|}{\multirow{2}{*}{\begin{tabular}[c]{@{}c@{}}\textbf{Event}\\\textbf{Num}\end{tabular}}} & \multicolumn{1}{c|}{\multirow{2}{*}{\begin{tabular}[c]{@{}c@{}}\textbf{Event}\\\textbf{Rate}\end{tabular}}} &\multicolumn{1}{c|}{\multirow{2}{*}{\begin{tabular}[c]{@{}c@{}}\textbf{Event}\\\textbf{Size}\end{tabular}}} \\
    & & & & & & \\ \hline
    DARPA-Cadets & 1 & 11 & 14 GB & 15 M& 16.87 eps &1013 Byte \\ \hline
    DARPA-Theia & 1 & 11 & 7.5 GB& 10 M& 11.25 eps &810 Byte\\ \hline
    DARPA-Trace & 1 & 11 & 62 GB&  72 M& 75.76 eps & 925 Byte\\ \hline
    Simulation & 5 & 12 & 23 GB & 50 M &  48.23 eps& 483 Byte\\\hline
    Production & 300+ & 5 & 16.85 GB & 17 M &  39.35 eps& 1064 Byte\\\hline
    \end{tabular}}
    \label{tab:ground-truth}
\end{table}

\noindent\textbf{\textsf{Datasets.}}
We used five datasets to evaluate the server-side cost of ProvDetector, UNICORN, and HOLMES. 
Among the three datasets, DARPA-Cadets, DARPA-Theia, and DARPA-Trace are open datasets from DARPA~\cite{DARPA-TC}. 
Production dataset is the real auditing data collected from a security company \anosec. 
Simulation dataset is an in-lab dataset we created for attack simulation. 
We provide more information of these datasets in Table~\ref{tab:memcost}.

Particularly, DARPA-Cadets contains three attacks during a 3-day-long period. The attacker exploited the vulnerabilities of an Nginx server and achieved C\&C by injecting the payload to an ``sshd'' process. The attacker repeated the attack 3 times. He failed the first 2 times but succeeded in the last one.
DARPA-Theia contains one attack. The attacker first exploited the Firefox backdoor to install executable files to disk. After two days, he used the vulnerability of a browser extension and resumed the prior attack by injecting the file that had been previously dropped to disk into the ``sshd'' process.
DARPA-Trace contains two attacks. The first one is a Firefox backdoor with the DRAKON malware in memory, and the second one is a Pine backdoor with a DRAKON dropper.

The Production dataset was collected by {\anosec}'s EDR deployed in the customers' network, which includes 300+ servers and working machines of employees of 10 real customers from \anosec, including schools, research institutes, factories, and healthcare providers. 
We monitored the customers for five days. 
We used the first three days (training period) to train the detection model and used the last two days (test period) for testing.

The Simulation dataset was collected from five hosts: one Ubuntu 20.04 server (U1), two Windows Server 2012 R2 Datacenters (S1,S2), one Windows Server 2019 Datacenter (S3),  and one Windows 10 desktop host (D1). We deployed Apache and PostgreSQL on Windows Servers and Nginx and PostgraSQL on Ubuntu 20.04 to simulate servers in the \anosec. We used the Windows 10 desktop to simulate the PCs used by the employees in the \anosec. 
The collected data had the same format as Sysdig for Linux and ETW for Windows.

\begin{table}[t]
    \setlength{\abovecaptionskip}{5pt}
    \scriptsize
    \centering
    \caption{Average number of graph nodes for the evaluation datasets and memory consumption of three \ac{pedr} systems}
    \resizebox{0.48\textwidth}{!}{\begin{tabular}{|l|r | r |r | r|}
    \hline
    \multicolumn{1}{|c|}{\multirow{2}*{\textbf{Dataset}}} & \multicolumn{1}{c|}{\multirow{2}{*}{\textbf{\# of \textbf{Graph Nodes}}}}  & \multicolumn{3}{c|}{\textbf{Memory (MB/host)}} \\ \cline{3-5}
    &  & \textbf{HOLMES}  & \textbf{ProvDetector} & \textbf{UNICORN}\\ \hline  
    DARPA-Cadets & 280W+ &   5683  & 10240 & 274     \\\hline
    DARPA-Theia & 125W+ &  3870 & 6574 &  242    \\ \hline
    DARPA-Trace & 325W+    & 9605 & - &  242      \\ \hline
    Simulation &  3W+ &     73 & 195 & 213     \\ \hline
    Production  &   5W+   & 84 & 240 & 219   \\ \hline
    \end{tabular}}
    \vspace{-0.2in}
    \label{tab:memcost}
\end{table}



\noindent\textbf{\textsf{Memory}}.
Table~\ref{tab:memcost} shows the memory consumption results. 
The memory consumed by HOLMES and ProvDetector was positively correlated with the data volume of the provenance graphs, which both exceeded the reference value ($<$20MB/host) by 1-2 orders of magnitude. 
For UNICORN, it had a relatively stable memory consumption because it used Parallel Sliding Windows (PSW) algorithm to analyze the whole provenance graph, which was independent of memory constraints. 
However, it exceeded the reference value by 11.9 times.
Therefore, none of these systems meet the requirement for the \textsf{Memory} factor and more memory consumption optimizations are needed for these systems.

\noindent\textbf{\textsf{Interpretation.}}
\begin{table}[t]
    \setlength{\abovecaptionskip}{5pt}
    \scriptsize
    \centering
    \caption{Interpretation cost (\# of graph nodes)}
    \resizebox{0.48\textwidth}{!}{\begin{tabular}{|l|r|r|r|}
    \hline
    \multicolumn{1}{|c|}{\textbf{Dataset}} &\multicolumn{1}{c|}{\textbf{ HOLMES}} &\multicolumn{1}{c|}{\textbf{ProvDetector}} &  \multicolumn{1}{c|}{\textbf{UNICORN}} \\ \hline
    DARPA-Cadets    &     173 & 15 & 154730                        \\ \hline
    DARPA-Theia &    73 & 8 & 522735                     \\ \hline
    DARPA-Trace   & 450 & - & 1454033                       \\ \hline
    Simulation  &     566 & 7 & 11587                     \\ \hline
    Production   & 81 & 5 & 17853                         \\ \hline
    \end{tabular}}
    \label{tab:graph-size}
\end{table}
Table~\ref{tab:graph-size} shows the result for the \textsf{Interpretation} factor. The provenance graphs generated by ProvDetecor can satisfy the reference value (< 50 nodes). 
HOLMES generates alarms within ten times larger than the reference value. 
Even worse, UNICORN reports the whole graph as an alarm and cannot pinpoint the concise location of attacks.
Thus, it generates too coarse-grained provenance graphs 3 to 4 orders of magnitude larger than the reference value, which is not practical in industry.

\noindent\textbf{\textsf{Alarm Triage.}}
\begin{table}[!t]
    \setlength{\abovecaptionskip}{5pt}
    \scriptsize
    \centering
    \caption{Average Alarm Number (alarms/host/day)}
    \resizebox{0.48\textwidth}{!}{\begin{tabular}{|l|r|r|r|}
    \hline
    \multicolumn{1}{|c|}{\textbf{Dataset}} &\multicolumn{1}{c|}{\textbf{ HOLMES}} &\multicolumn{1}{c|}{\textbf{ProvDetector}} &  \multicolumn{1}{c|}{\textbf{UNICORN}} \\ \hline
    DARPA-Cadets    & 21 & 90 & 0.3                        \\ \hline
    DARPA-Theia &    36.7 & 90 & 0.1                     \\ \hline
    DARPA-Trace   & 13.9 & - & 0.45                       \\ \hline
    Simulation  & 2.3 & 23 & 0.09                     \\ \hline
    Production & 12.1  & 56.3 & 0.13                         \\ \hline
    \end{tabular}}
    \label{tab:alert-number}
\end{table}
As shown in Table~\ref{tab:alert-number}, only UNICORN can roughly satisfy the reference value (<0.1 alarms/host/day). 
HOLMES and ProvDetector still need to reduce the number of alarms by more than 2 orders of magnitude to meet the reference value. 
Specifically, since improving precision can reduce the number of alarms per host per day (See Section~\ref{sec:laborcost}), HOLMES and ProvDetector will need to improve their precision significantly.

%% file: tex/findings.tex
\section{Findings of Our Study}
\label{sec:findings}
In this section, we summarize the key findings of our study and answer the three research questions. 
In particular, we address RQ1 and RQ2 based on the results of the interviews, and address RQ3 based on the results of all the four studies.

\subsection{RQ 1: Effectiveness of \ac{pedr}}
In our interviews, we found that the managers in the industry all agreed that \ac{pedr} was more effective than conventional \ac{edr} systems due to better interpretability.  
They all showed great interest in \ac{pedr} systems and agreed that \ac{pedr} systems had great potential to replace conventional \ac{edr} systems. 

\noindent\textbf{Replacing \ac{edr} Systems}.
As shown in Table~\ref{tab:Interview_backgroud}, 4 out of the 10 managers have adopted \ac{pedr} systems to replace the conventional \ac{edr} systems in their products or environments. 
For instance, $E1$ said: ``We use provenance analysis techniques for attack investigation. The \ac{pedr} takes the alarm event as the starting point and generates a limited provenance graph through causal analysis for manual confirmation. The contextual information contained in the provenance graph greatly improves the efficiency of attack investigation.'' 
$E7$, the developer of a \ac{pedr} system, also said: ``Our customers are interested in the improvement of attack detection and investigation brought by provenance analysis techniques, so we decide to focus on \ac{pedr} systems.'' 


Even the managers who were not using \ac{pedr} systems showed great interest in \ac{pedr} systems. 
They had not adopted \ac{pedr} systems yet due to the higher cost.
For example, $E8$ said: ``We attempted to detect attacks using provenance graphs on a customer with 1,200 hosts. However, 800MB of memory is required to detect the attack for the provenance graph data of only one host, and the experimental server runs out of memory after running only 40 host data. We cannot afford the memory cost. Nevertheless, we still hope to find a feasible method.''

\noindent\textbf{Interpretability}.
The managers agreed that it was straightforward to interpret the results of \ac{pedr} systems. 
Surprisingly, even the basic provenance graphs that consist of low-level system audit events are easy to interpret for security analysts as long as they are concise. 
For example, $E2$ says: ``An analyst's ability to translate alarm semantics is related to his experience, and most matured analysts seldom encounter this problem. Even inexperienced newbies can understand the provenance graphs by taking a quick training.''. 
On average, a novice analyst can understand most provenance graphs by taking a 7-14 days training session, as mentioned by $E1$ and $E6$. 
Lastly, the managers all agreed that existing techniques that abstract the basic provenance graphs to more intuitive levels, such as technique and tactic levels~\cite{TTP,bahrami2019cyber,8258514}, can potentially improve the interpretability of provenance graphs.

\begin{tcolorbox}[title=Answer to RQ 1, left=2pt, right=2pt, top=2pt, bottom=2pt]
The industry acknowledges that \ac{pedr} systems are superior to conventional \ac{edr}  systems due to better interpretability. 
Experienced security analysts can easily understand basic provenance graphs that consist of low-level system audit events, and companies have designed training sessions in provenance analysis for training novice analysts.
\end{tcolorbox}

\subsection{RQ 2: Adoption Bottlenecks}
\label{sec:rq2}
According to the interview results, \textit{the primary bottleneck for the industry to adopt \ac{pedr} systems is the cost instead of the performance}. 
In fact, only two managers ($E2$ and $E3$) considered detection accuracy as a decision factor, and they still considered it as an optional factor and ranked it after other factors. 
The major reason is that these managers already have mature processes in working with existing \ac{edr} systems that generate lots of false positives, and \ac{pedr} systems generally have better detection accuracy than \ac{edr} systems.

Through discussions with these managers, we realized that the decision process of an industrial manager to adopt a \ac{pedr} system, or an \ac{edr} in general, was to minimize the potential loss of successful attacks and the cost of running an \ac{edr} system. 
Formally, the managers aim to minimize the $overall\_cost$,
where $overall\_cost = loss_e + op\_cost$, $loss_e$ is the expected loss, and $op\_cost$ is the operating cost that consists of computing cost and labor cost (Section~\ref{subsec:interview_result}).
Here, $loss_e$ is considered as a constant because it is not observable in practice. 
Thus, when a manager was evaluating an \ac{edr} system, he first tested whether the \ac{edr} system could detect attacks in a testing environment with sufficient accuracy. As long as the detection recall exceeds a certain threshold, the manager replaces $loss_e$ with a constant. Furthermore, almost all existing \ac{pedr} systems can achieve higher recalls than their thresholds, as mentioned by the managers.  
Therefore, the managers only considered the operating cost as the primary bottleneck of a \ac{pedr} system. 
This also explains why none of them chose recall as one of decision factors. 

\begin{tcolorbox}[title=Answer to RQ2, left=2pt, right=2pt, top=2pt, bottom=2pt]
The operating cost, which consists of the four-must factors: \textsf{Memory}, \textsf{Client-Side Overhead}, \textsf{Interpretation}, and \textsf{Alarm Triage}, is the primary bottleneck for the industry to adopt an \ac{edr}/\ac{pedr} system. 
\end{tcolorbox}

\subsection{RQ3: Gaps Between Industry and Academia}
\label{sec:gapbetweenindustryacademia}
According to the results of all the four studies, we find that there are three important gaps between the \ac{pedr} techniques proposed by academia and the expectations of the industry.

\noindent\textbf{Gap 1: Overlooking \textsf{Client-Side Overhead}}.
Although the industry considers the \textsf{Client-Side Overhead} as one of the most important factors for adopting \ac{pedr} systems, academia often overlooks it. 
Based on our interviews, 8 out of 10 managers identified the client-side overhead as the most important decision factor.
However, all the 20 surveyed papers, except for RTAG, did not evaluate the client-side overhead of their approaches. 
Worse still, by investigating existing provenance collectors in academia and industry, we found that there were no comprehensive evaluations on the client-side overhead of these collectors, even though the most popular commercial provenance collectors (Auditd, Sysdig, LTTng, and ETW) shown in Table~\ref{tab:collectors}.
Through our literature review (Section~\ref{sec:LiteratureSurvey}) and measurement study (Section~\ref{sec:measurementserverside}), we found that existing provenance collectors could not satisfy the reference value of runtime overhead ($<$3\%).

\noindent\textbf{Gap 2: Imbalance between \textsf{Alarm Triage} and \textsf{Interpretation}}.
\textsf{Alarm triage}, and \textsf{Interpretation} are two must-meet factors for the \ac{pedr} systems.  
Our measurement study shows that none of the existing \ac{pedr} systems meet both of these factors. 
With deeper investigation, we realize that these \ac{pedr} systems implicitly sacrifice one factor to enhance the other. 
Consider the three representative systems (UNICON, HOLMES, and ProvDetector) in Table~\ref{tab:alert-number}. 
According to our study, UNICORN has a satisfying alarm triage cost. 
However, this comes with the interpretation cost of several orders of magnitudes higher than the other two systems. 
This is because UNICORN projects provenance graphs into embedding vectors, which improves detection accuracy,
but the projection also prevents UNICORN from pruning irrelevant events from the provenance graphs, leading to huge graphs (millions of nodes). 
On the contrary, HOLMES and ProveDetector detect anomaly paths in provenance graphs, generating much smaller graphs (low investigation costs) but resulting in much higher false positives (dozens on average).

\noindent\textbf{Gap 3: Excessive Server-Side Memory Consumption}.
\textsf{Memory} is a must-meet factor for adopting \ac{pedr} systems.
But our literature survey shows that academia has not paid attention to server-side memory consumption, and our measurement study shows that existing \ac{pedr} systems cannot satisfy the reference value ($<$20MB/host).
The root cause for such intolerable memory consumption is that these systems cache all the provenance data in the memory, such as HOLMES and ProveDetector.
Therefore, these systems cannot scale to monitor large clusters of hosts.
For example, ProveDetector failed to conduct detection on the DARPA-Trace dataset due to memory explosion. 
Unlike these two systems, UNICORN adopts a stream-based processing approach that uses a sliding window to cache only the most recent provenance data.
Even so, its memory consumption is still 200MB/host, which is about 10$\times$ of the reference value (20MB/host).

\begin{tcolorbox}[title=Answer to RQ 3 (derived from all four-part studies), left=2pt, right=2pt, top=2pt, bottom=2pt]
There exist three important gaps (overlooking client-side overhead, the imbalance between alarm triage cost and interpretation cost, and excessive server-side memory consumption) between the academic research and the industry expectations.
\end{tcolorbox}

%% file: tex/discusion.tex
\section{Discussion}

\subsection{Study Limitations}
The limited number of participants in our one-to-one interview may harm the generalizability of our study. To address this threat, we recruited participants from different top IT companies, including both customers and providers of \ac{pedr} systems. Further, we followed up the interviews with an online questionnaire that expanded the scope of the participants.
We also strictly followed the principles in \textit{Qualitative Interview Design}~\cite{turner2022qualitative,mann2016research} and \textit{How to Design and Frame a Questionnaire}~\cite{farooq2018design} when conducting the interviews and the follow-up questionnaires.
Further, inaccurate implementation and inappropriate parameter configurations of the chosen \ac{pedr} system may also harm the validity of our study.
To mitigate this threat, we used the original implementations if they were available or strictly followed the paper descriptions to implement and configure the systems (e.g., ProvDetector and HOLMEs).
We also share the systems~\cite{EDREmpiricalStudy} and the evaluation datasets with the community for subsequent reproducible research.

\subsection{Implications}
Our study findings (Section~\ref{sec:findings}) identify potential areas to improve \ac{pedr} techniques. 
We summarize the study implications with the focus on filling the important gaps as follows. 

\noindent\textbf{Adopting Data Reduction for Gap 1:} 
To date, client-side overhead has received less attention than others and more efforts are desired to optimize the runtime overhead of collectors. 
Recent studies on provenance data reduction~\cite{reduction, reduction2, reduction3} show that there are a large number of repeated and similar logs in the collected logs, which waste a lot of memory on the client side. 
Thus, a promising approach is to integrate causality-preserving reduction~\cite{reduction} and other data reduction techniques to provenance collectors to greatly reduce the volume of log data. 
However, existing data reduction techniques are mainly designed to run on the server side, and complex compression algorithms are too expensive to be directly applied to the client's collector. 
For example, NodeMerge~\cite{reduction2} requires 928.61MB of memory,
and efficient collectors pursue smaller overhead rather than data compression ratio. 
Therefore, we can develop a lightweight collection and filtering framework to reduce the collection of irrelevant log data through lightweight computation such as heuristic rules on identifying temp files~\cite{loggc} or deprioritizing chronicle maintenance processes.

\noindent\textbf{Integrating Alarm Filtering for Gap 2:} 
Due to the lack of industry insights, existing work mainly focuses on how to reduce the number of alarms and ignores the size of the alarm graph.
In addition, many of the key papers~\cite{hassan2020rapsheetl, hassan2019nodoze, HOLMES} related to alarm filtering mostly adopt a single filtering method such as alarm correlation or alarm ranking, and the filtering effect on large-scale clusters is insufficient. 
For example, NoDoze~\cite{hassan2019nodoze} is an alarm ranking technique that assigns an anomaly score based on the frequency to combat threat alarm fatigue produced by the rule-based host IDPS. 
The filtering effectiveness of NoDoze is only around 84\%. 
If it is applied to the production data set of the HOLMES in Section~\ref{sec:measurementserverside}, there are still 1.94 alarms/host/day, which is far from the industry reference value (< 0.1 alarms/host/day).
To reduce the amount of alarms, we can adopt a systematic alarm filtering method, which can integrate alarm aggregation, correlation, and ranking methods, reaching the desirable alarm level. 
At the same time, for those systems that generate a large-scale alarm graph, we can design an alarm graph clipping algorithm to identify and delete irrelevant nodes and edges in the alarm graph, so as to control the graph size to a reasonable range.

\noindent\textbf{Distributing Server Workload and Archiving Events for Gap 3:} 
The key papers discussed in our study all adopt a centralized architecture, which uploads the full amount of logs to the server, and then builds a complete provenance graph for complex graph clipping and matching calculations to detect attacks. 
However, building and maintaining provenance graphs require a lot of memory, 
and yet a large portion of nodes and edges in the provenance graph is irrelevant to actual attacks~\cite{hassan2020rapsheetl}, wasting a lot of memory. 
Thus, a promising solution is to adopt a distributed architecture to utilize client computing if clients have spared computing capacity to reduce server memory burdens. 
For example, we can design a lightweight filtering algorithm on the client side to identify suspicious events, and only upload information related to suspected attack events to the server. 
Unlike the centralized architecture, which needs to reconstruct and maintain provenance graphs and perform complex computations on these graphs, a distributed architecture only processes localized data of suspicious events, and the required memory is greatly reduced. 
Furthermore, we can design an algorithm to periodically evict the events cached in the memory to the hard disk during attack detection and fetch the associated data from the disk when it is needed during attack investigation.

\eat{
\noindent\textbf{Future Research Directions}.
Based on our findings, there exsit three improtant gaps between academia and industry: \textit{overlooking client-side over-head},\textit{imbalance between alarm triage cost and interpretation cost}, \textit{excessive server-side memory consumption}. The challenges of bridging these gap are as follows:
\noindent\textbf{Challenge 1:} How to collect complete provenance data under the limitation of client memory cost?
\noindent\textbf{Challenge 2:} How to balance the number of alarms and the detection granularity?
\noindent\textbf{Challenge 3:} How to use the provenance graph for attack detection and investigation under the memory limit of the server?
In response to the above challenges, we propose future research directions:
\begin{itemize}
     \item For Challenge 1, existing provenance collectors also suffer the event-dropping vulnerability: they drop events massively when the workload of a system is high. This vulnerability is particularlly problematic because it allows attackers to disable provenance collectors without the root privilege by generating massive meaningless provenance events in a short period of time. For example, the owner of Sysdig discovered a flooding attack on their system provenance collector (CVE 2019-8339) ~\cite{sysdigcve}. The alternative solution is to elastically allocate sufficient resources for the collector to ensure that all provenance data can be processed in time. However, this strategy will degrade the performance of the whole system. How to Ensure the integrity of provenance data as well as the performance of the applications is a great challenge to both academia and industry, raising an important issue to be addressed in the future.
     \item For Challenge 2, to reduce the amount of alarms, we can adopts a systematic alarm filtering method, which can integrate alarm aggregation, correlation, and scoring methods, reaching the desirable alarm level. To control the detection granularity in a readable.
     \item For Challenge 3, we can distribute part of the pressure of the central detection server to each client, which can be called distributed \ac{pedr}. For example, the clients only transmit valuable information to the server and discard a large number of attack-irrelevant events, so that it can reduce the burden of the server side. In addition, we can design an algorithm to periodically evict the events cached in the memory to the hard disk while detection and fetch associated data from the disk during the investigation. 
 \end{itemize}
}



\eat{
总结
目前学术界的PEDR少有同时满足四个必要条件，导致了P-EDR在工业界使用率低。仅有的四家P-EDR厂商，也仅仅是将Provenance graph用于溯源调查中，生成简单的溯源图便于分析师确认。学术界使用provenance graph进行攻击检测，复杂的图匹配计算，还没有在工业产品中得到很好的应用，高昂的内存成本是主要的障碍。
主要的gap在于：gap 1，内存消耗        gap2，服务端内存消耗     gap3  告警疲劳
主要的挑战在于：（1）如何在客户端内存成本限制下，采集和使用provenance data；
（2）如何在服务端的成本限制下，使用provenance graph进行攻击检测和溯源
（3）如何在图节点范围内，控制告警数量？
针对上述挑战，我们提出未来的研究方向：挑战1研究方向：智能采集，通过减少无效重复的事件，减少日志数量，降低处理成本；挑战2研究方向，使用分布式架构，合理的利用客户端和服务端的；有限图计算，通过在有限范围溯源图上进行计算，减少图构建和图匹配的计算量；针对挑战3，采用系统性的告警过滤方法，可以将告警聚合，关联，评分方法集成起来，达到告警水平。

}

\eat{

One way to obtain 

How to obtain accurate industry information on the status of \ac{pedr} is important for this study. 
It is another better way for researchers to enter the company or do an internship. However, this approach is difficult because different companies have different policies.
1.采访和问卷 response accurate
如何获取准确的工业界对P-EDR现状信息对本研究很重要。
研究人员进驻公司或者实习是另外一种比较好的方法。
然而，这种方法很困难，因为不同公司的进驻政策不同。为了保障从采访和问卷中获得的信息的准确性，我们做了以下措施。1.充足时间来保障受访者的意愿。我们提前跟采访中的所有技术经理预约时间，并且准备好采访相关材料。我们控制问卷的长度在10分钟以内。2.隐私保护让受访者真实的回答。所有受访者都会被提前告知，采访或问卷不会涉及个人因素，数据不会暴露个人隐私。如果涉及到敏感信息的问题是可选的。

2.empirical study
典型PEDR系统的实现和实验参数配置是empirical study的结果准确性的重要因素。
为了缓解这个因素的影响，我们尽我们最大努力按照paper中的描述实现了homes和provdetector，并且采用paper中的最佳参数配置来进行实验。我们将两个工具共享到社区以便后续复现研究。
}

%% file: tex/related.tex
\section{Related Work}
Researchers have shown great interest in understanding the challenges and opportunities of \ac{pedr} systems. Han et.al.~\cite{han2018provenance} summarized the opportunities
and challenges associated with \ac{pedr} and provide insights based on their research experience in this area. 
Li et.al.~\cite{li2021threat} conducted a literature review on existing \ac{pedr}s in academia. The most recent measurement study conducted by Inam et.al.~\cite{SoK-History} summarizes \ac{pedr} related techniques published in the top-tier system and security conferences and builds taxonomy based on the system auditing pipeline. Alahmadi et.al.~\cite{alahmadi202299} carried out a qualitative study of conventional \ac{soc} analysts' perspectives on security alarms through an online survey and semi-structured interviews. Yet, none of the existing papers have studied the effectiveness and bottlenecks of \ac{pedr} systems from the perspective of the industry. Note that, the most well-known \ac{pedr} systems are introduced in Section~\ref{sec:LiteratureSurvey}.

%% file: tex/conclusion.tex
\section{Conclusion}
In this paper, we conduct the first set of systematic studies on the effectiveness and the bottlenecks of existing \ac{pedr} systems from the industrial perspective. We also conduct a literature survey and a measurement study to identify the gaps between the techniques developed in academia and the expectations of the industry. Our study shows that the industry believes that \ac{pedr} systems are superior to convention \ac{edr} systems. However, the industry is also concerned about the operating cost of \ac{pedr} systems. We further identify three gaps between academia and the industry. Particularly, we find the academia (1) overlooks the client-side overhead of \ac{pedr} systems, (2) fails to balance alarm triage and interpretation, and (3) needs to significantly reduce the server-side memory consumption for \ac{pedr} systems. Taken together, we expect these findings to help improve researchers’ understanding of the expectations of \ac{pedr} systems from the industry. 

%% file: tex/appendix.tex
\appendix
\section{Background of Questionnaire Participants}
\label{appendix:questionnaire}
\begin{figure}[]
    \setlength{\abovecaptionskip}{5pt}
    \centering
    \subfigure[Organization type]{
    \includegraphics[width=0.32\textwidth]{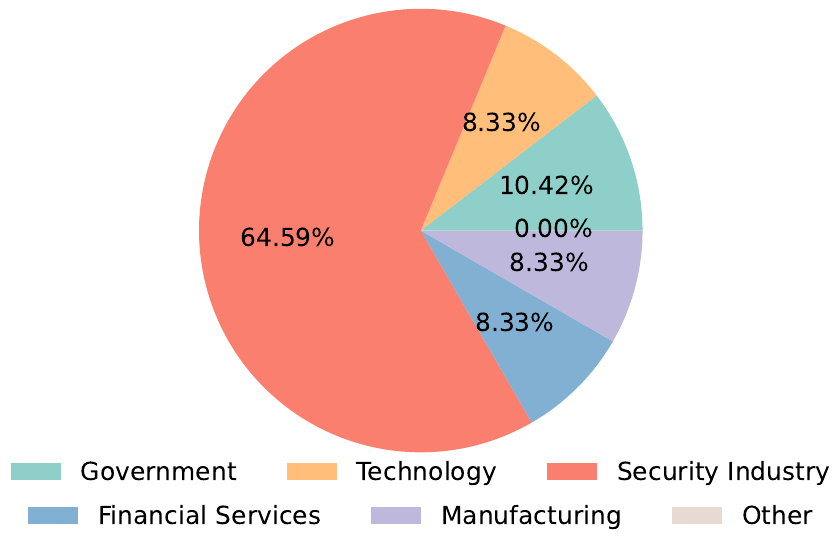}
    \label{fig:ASGf1}}

    \subfigure[Position in the company]{
    \includegraphics[width=0.32\textwidth]{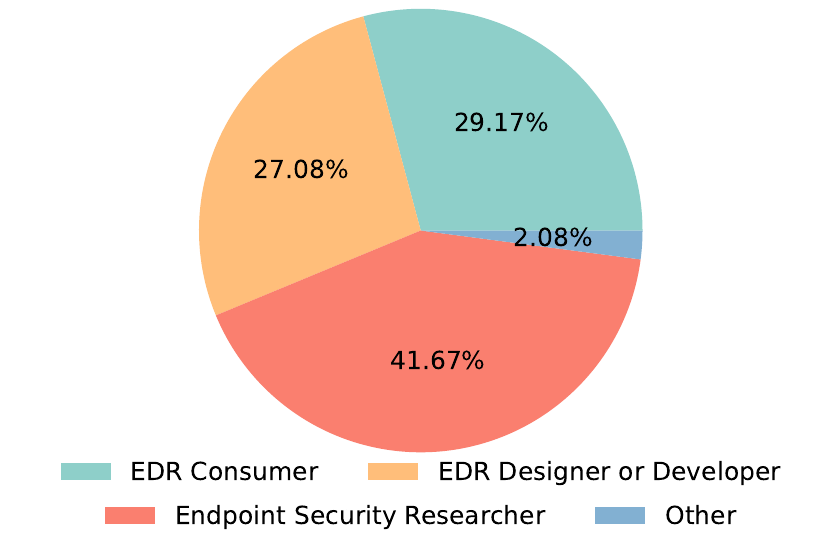}
    \label{fig:kvalue}}

    \subfigure[APT combating experiences]{
    \includegraphics[width=0.32\textwidth]{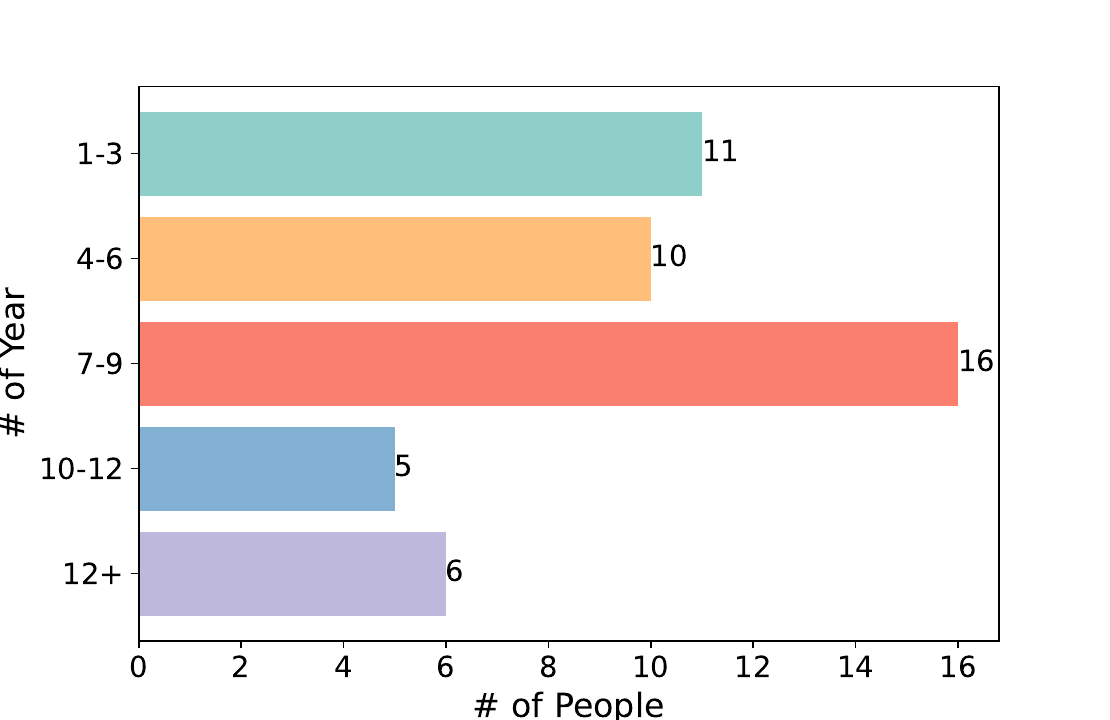}
    \label{fig:ASGSize}}   
    \caption{Participants of our online questionnaire}
    \label{fig:participants}
\end{figure}

Figure~\ref{fig:participants} shows the type of organizations, position in company, \ac{apt} combating experiences of participants in our questionnaire study.

\eat{
\begin{figure}[h]
    \setlength{\abovecaptionskip}{5pt}
    \includegraphics[width=0.48\textwidth]{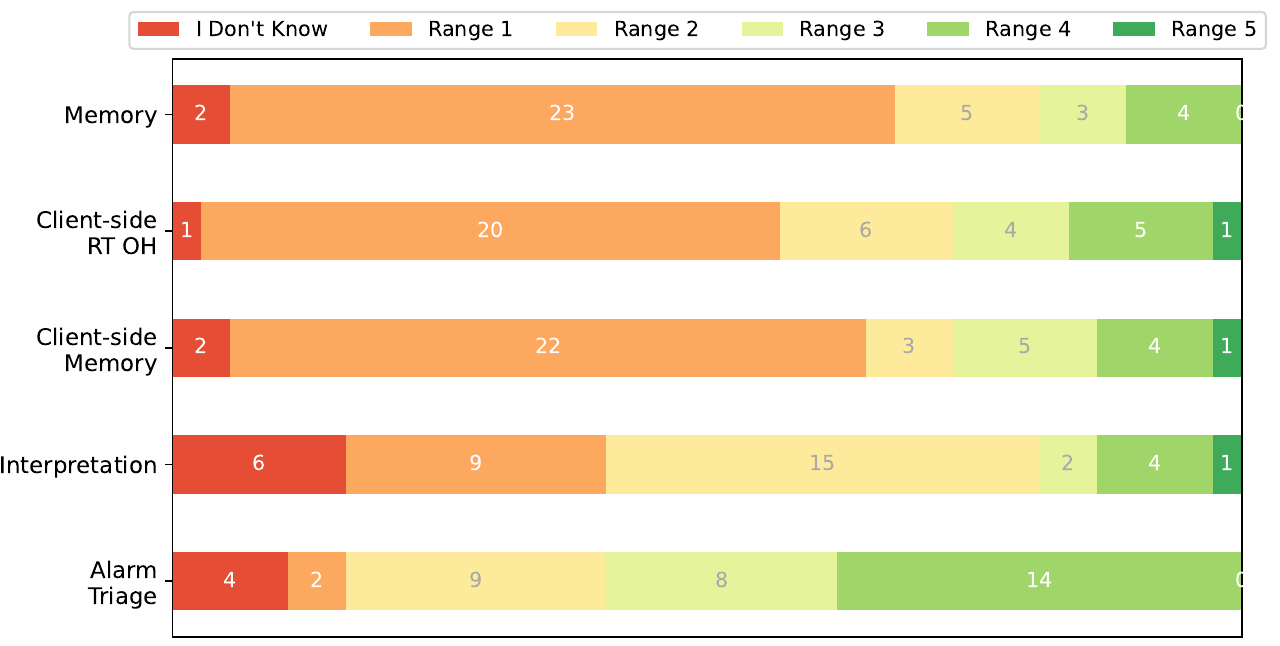}
    \caption{Metric Results of Our Questionnaire Study.}
    \label{fig:questionnaire-results}
\end{figure}
}

\section{Online Questionnaire}
\label{appendix:online}
Our team has been working on provenance-based endpoint detection and response tools (\ac{pedr}) for Advanced Persistent Threats (APT) detection for many years. We are currently working on a 10-minutes questionnaire for understanding the industry's expectations about P-EDR systems. The finding may inspire us to design better \ac{pedr} systems. 
This work was approved by our institution and we strictly follow our institution's research data management policy. Your personal privacy information will be carefully processed and your thoughts will be accurately reflected in our study. Thanks for your participation!

\begin{enumerate}
    \item What is your job type?
    \begin{itemize}
        \item EDR Consumer
        \item EDR Designer or Developer
        \item Endpoint Security Researcher
        \item Others
    \end{itemize}
    \item How many years of experience do you have in your role?
    \begin{itemize}
        \item 1 - 3	
        \item 4 - 6
        \item 7 - 9
        \item 10 - 12
        \item 12+
    \end{itemize}
    \item How would you rate your level of expertise in endpoint security monitoring?
    \begin{itemize}
        \item Very low
        \item Low
        \item Medium
        \item High
        \item Very high
    \end{itemize}
    \item What is the type of organization you work in?    
    \begin{itemize}
        \item Government
        \item Technology
        \item Security Industry
        \item Financial Services
        \item Manufacturing
        \item Others
    \end{itemize}
    \item     How many endpoints need to be protected in the \ac{soc} you work in?
    \begin{itemize}
        \item 0 - 1000 hosts	
        \item 1001 - 10,000 hosts	
        \item 10,001 - 100,000 hosts	
        \item 100,001 - 1,000,000 hosts	
        \item more than 1,000,000 host
        \item I don't know
    \end{itemize}
    \item What is the minimum amount of memory required on the machine where your \ac{edr} server is installed and running?
    \begin{itemize}
        \item 8 - 16GB		
        \item 17 - 32GB		
        \item 33 - 64GB	
        \item 65 - 128GB	
        \item more than 128GB 
        \item I don't know
    \end{itemize}
    \item  What is the average maximum amount of the host RAM memory that the \ac{edr} local agent runtime can occupy?   
    \begin{itemize}
        \item < 100MB/host	
        \item < 150MB/host	
        \item < 200MB/host
        \item < 250MB/host
        \item more than 250MB/host
        \item I don't know
    \end{itemize}
    \item     What is the average maximum percentage of the host CPU that the edr local agent runtime can occupy?
    \begin{itemize}
        \item 1 - 3\%	
        \item 4 - 5\%	
        \item 6 - 8\%	
        \item 9 - 10\%	
        \item more than 10\%
        \item I don't know
    \end{itemize}
    \item     How many security analysts are responsible for the above hosts?
    \begin{itemize}
        \item 1
        \item 2
        \item 3 - 4
        \item 5 - 10
        \item more than 10
        \item I don't know
    \end{itemize}
    \item How many \ac{edr} alarms can a security analyst investigate on average per day (calculated on an eight-hour work schedule)?
    \begin{itemize}
        \item < 100 alarms/day/person
        \item < 150 alarms/day/person	
        \item < 200 alarms/day/person	
        \item < 250 alarms/day/person	
        \item more than 250 alarms/day/person
        \item I don't know
    \end{itemize}
    \item For graph-based \ac{apt} detection device, how many nodes does the detected graph need to be controlled within?
    \begin{itemize}
        \item < 30 nodes	
        \item < 50 nodes	
        \item < 70 nodes	
        \item < 100 nodes	
        \item more than 100 nodes
        \item I don't know
    \end{itemize}
\end{enumerate}